\documentclass[12pt]{spieman}  
\usepackage{amsmath,amsfonts,amssymb}
\usepackage{graphicx}
\usepackage{setspace}
\usepackage{tocloft}
\usepackage{hyperref}
\usepackage[switch,columnwise]{lineno}
\usepackage{aasmacrospk}
\DeclareGraphicsExtensions{.pdf,.png,.jpg,.eps}

\newcommand{\ie}{i.e.,\ }
\newcommand{\etal}{et~al.\ }

\newcommand{\eg}{e.g.,\ }

\def\ie{{\it i.e.,}\ }
\def\eg{{\it e.g.,\ }}
\def\simgt{\ {\raise-.5ex\hbox{$\buildrel>\over\sim$}}\ }
\def\simlt{\ {\raise-.5ex\hbox{$\buildrel<\over\sim$}}\ }

\def \h70{{h_{70}}}

\title{Enabling Discoveries: Thirty Years of Advanced Technologies and Instrumentation at the National Science Foundation} 

\author[a,b,*]{Peter Kurczynski}
\author[c]{and Sta\v{s}a Milojevi\'{c}}
\affil[a]{National Science Foundation, 2415 Eisenhower Ave, Alexandria , Virginia, USA}
\affil[b]{Rutgers, the State University of New Jersey, Piscataway, New Jersey, USA}
\affil[c]{Center for Complex Networks and Systems Research, Luddy School of Informatics, Computing, and Engineering, Indiana University, Bloomington, USA}

\cftpagenumbersoff{figure}
\cftpagenumbersoff{table} 
\begin{document} 
\maketitle 
\begin{abstract}
Over its more than thirty-year history, the Advanced Technologies and Instrumentation (ATI) program has provided grants to support technology development and instrumentation for ground-based astronomy. Through a combination of automated literature assessment and in-depth literature review, we present a survey of ATI-funded research and an assessment of its impact on astronomy and society.  Award acknowledgement and literature citation statistics for ATI are comparable to a comparison astronomy grant program that does not support technology development. Citation statistics for both NSF-funded programs exceed those of the general astronomical literature. Numerous examples demonstrate the significant, long term impact of ATI-supported research on astronomy. As part of this impact, ATI grants have provided many early career researchers the opportunity to gain critical professional experience. However, technology development unfolds over a time period that is longer than an individual grant.  A longitudinal perspective shows that investments in technology and instrumentation have lead to extraordinary scientific progress.
\end{abstract}


{\noindent \footnotesize\textbf{*}Send correspondence to P.K. at \linkable{pkurczynski@physics.rutgers.edu}, Telephone 703-292-7248.  The views expressed here are solely those of the authors and do not reflect the views of the National Science Foundation.}

\begin{spacing}{2}   

\section{Introduction}
\label{IntroductionSection}

Ever since the dawn of modern astronomy, better observations have enabled better understanding of the universe. For example, precise naked-eye observations of the planets by Tycho Brahe circa 1609, accurate to about 0.1 degrees, and calculations by Johannes Kepler established that the Earth revolved around the Sun and that planetary orbits were elliptical\cite{Kuhn1957}. In modern times, the National Science Foundation Laser Interferometer Gravitational-Wave Observatory (NSF-LIGO) measured the movement of test masses to less than an atomic diameter. This extraordinary precision enabled the discovery of gravitational waves and gave birth to a new field of gravitational wave astronomy.

Technology and instrumentation opens new possibilities for accurate and precise observations.  Today's development of new technologies and instrumentation for astronomy is supported in part by federal funding agencies.  The National Science Foundation supports this development  through the Advanced Technologies and Instrumentation (ATI) program within the Division of Astronomical Sciences. The ATI program fulfills a vital and unique role. It is the only program supporting ground-based astronomy that is specifically oriented toward developing advanced technologies that may be too risky or otherwise unsuitable for mid-scale or larger investment. Furthermore, ground-based applications are often the best way to develop technology or achieve certain science goals.

It is not clear how best to assess the impact of technology and instrumentation for astronomy. While it may be obvious that without telescopes and detectors, modern astronomy would not exist, in an environment of constrained budgets and competing funding priorities, the relative importance of new technology and instrumentation cannot be taken for granted. With this spirit in mind, this paper presents an historical overview of the ATI program with the goal of addressing the impact of ATI supported research.

The effects of a particular grant program can be assessed through analysis of the scientific literature.  Research publications that follow directly from an award, which investigators are required to acknowledge in their publications, document the immediate impact of funding.  However, additional indirect effects may be more far-reaching, though harder to quantify.  A new technological solution to a particular problem may enable a wide range of scientific investigations, spur additional innovation in the field or comprise a critical role in a larger initiative.  These impacts may not be fully apparent until years or even decades after an initial award.   Also important are the education and training opportunities that a research program provides.  The tools and methods of astronomy are increasingly specialized; opportunities to gain mastery of these methods are rare and precious.  Such indirect impacts of awards will be illustrated through particular examples over the history of the program.  

Section 2 describes the ATI program and places it in context of other NSF programs.  Section 3 presents the results of automated literature citation analysis and comparison to a ``pure science" program that does not emphasize technology development.  Section 4 presents a narrative history of the program and some of the illustrative awards and their resulting impact.  The information here is obtained from the published literature as well as from conversations with individual investigators and experts in the field.  Section 5 summarizes the results and presents conclusions.

\section{Advanced Technologies and Instrumentation}
\label{ATISection}
The National Science Foundation (NSF) was established in 1950 as a federal agency for fostering civilian scientific research\cite{Mazuzan1992}. NSF is funded through Congressional appropriations that amounted to approximately \$8B in Fiscal Year (FY) 2019; it has a portfolio that spans the entire range of research and education in science and engineering.  The Division of Astronomical Sciences has an annual budget of approximately \$250M that supports ground-based astronomical facilities as well as individual investigator grants.  
   \begin{figure}
   \begin{center}
   \begin{tabular}{c}
   \includegraphics[height=7cm]{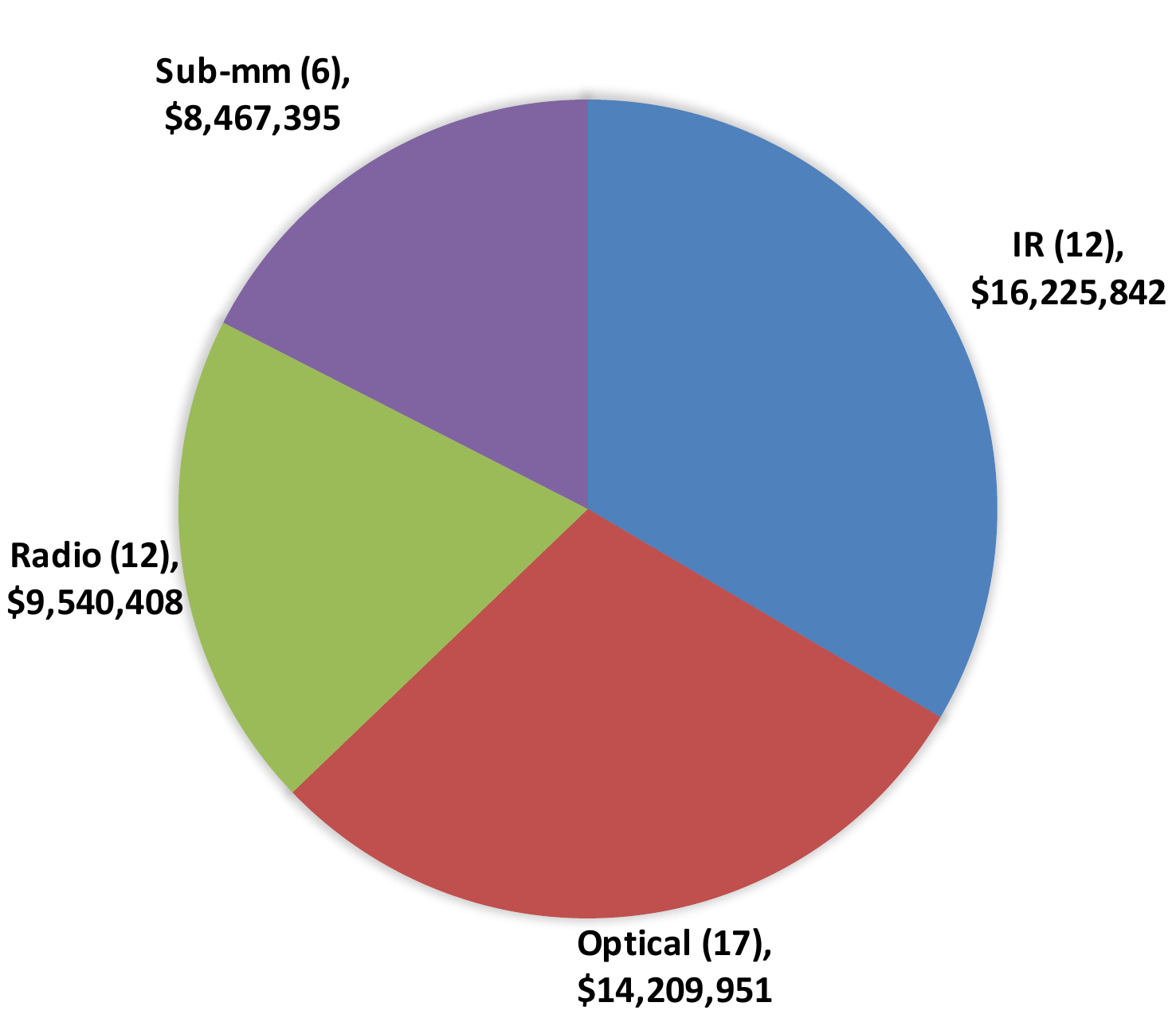}
   \includegraphics[height=7cm]{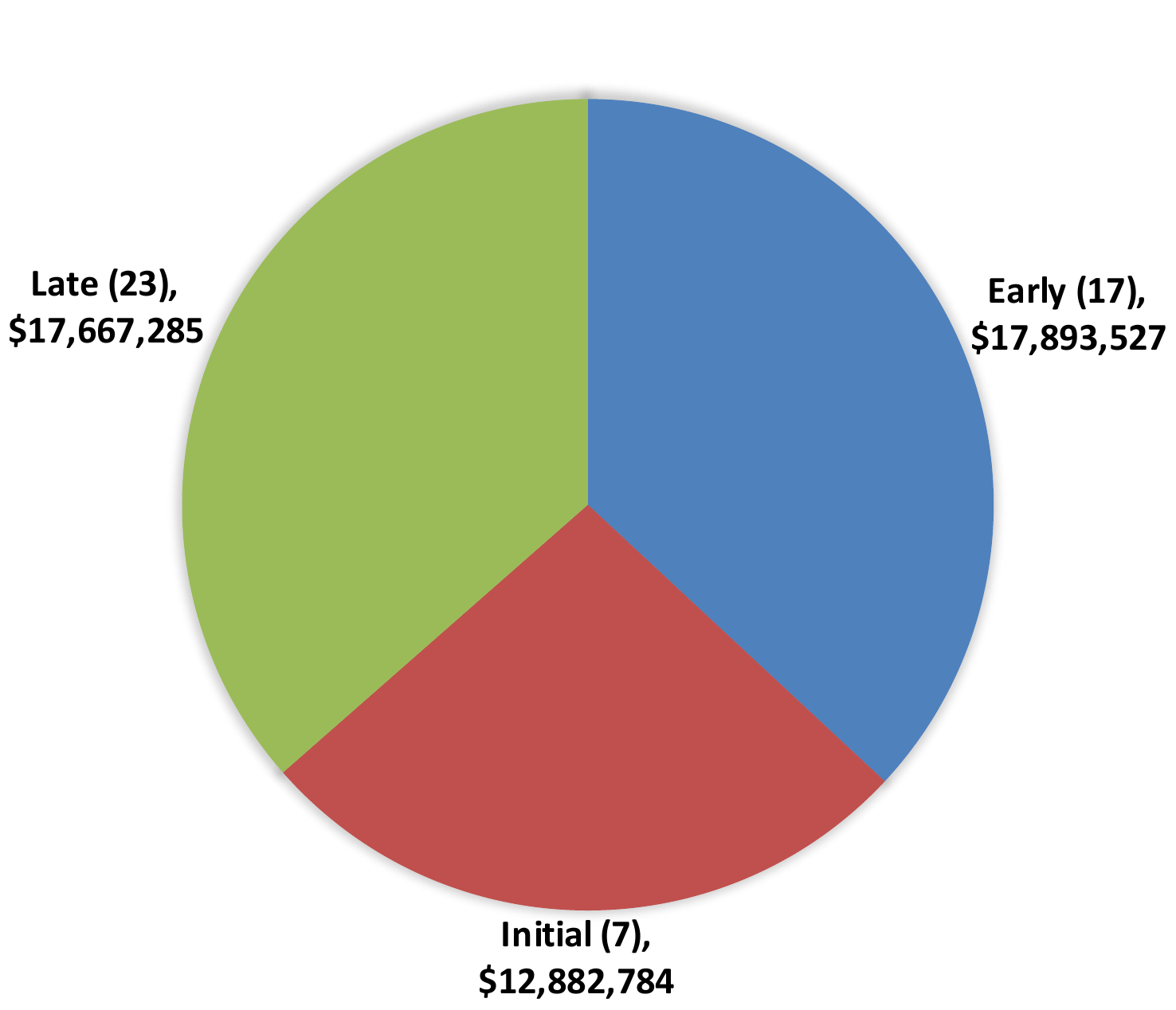}
   \end{tabular}
   \end{center}
   \caption[ActiveAwardsByWavebandAndTechDevp] 
   { \label{fig: ActiveAwardsByWavebandAndTechDevp} 
Summary of active ATI awards by waveband {\bf (left panel)} and technology development phase {\bf (right panel)}.  Size of each pie segment indicates the amount awarded to date (also shown in dollars); the corresponding number of awards are indicated in parentheses.  Awards are distributed across radio, sub-millimeter, IR and optical wavebands.  The active portfolio is broadly diversified among initial concepts and more mature technologies, as discussed in the text.}
   \end{figure} 

The bulk of astronomical research that is supported by individual investigator programs are included within the Astronomy and Astrophysics Grants (AAG) program. Grants are also awarded to support graduate student and post doctoral fellowships, undergraduate training, prestigious CAREER and other awards.  The Mid-Scale Innovations Program (MSIP) and the smaller ATI program each support new instrumentation and technology development for astronomy.  Deployment of existing technology for astronomical applications is supported through the Foundation-wide Major Research and Instrumentation (MRI) program.  The ATI and MSIP programs focus on technology development and/or instrumentation in support of specific scientific objectives.  ATI has supported research into adaptive optics, high resolution and multi-object spectroscopy, optical interferometry and synoptic surveys, to name just a few. Basic information on all NSF awards are publicly available. Such information includes the NSF award ID (a seven digit number beginning with the two-digit year of the award; \eg award 8911701 was awarded in FY 1989), principal investigator name and organization (institutional affiliation), title and abstract, program name, start and end dates and awarded amount to date.  These  public data for the ATI program from 1987 - 2016 form the basis for the analyses presented here.  Hereafter, where specific awards are mentioned, they will be referred by their NSF award ID followed by the principal investigator last name, \eg 8911701/Angel. 
   \begin{figure}
   \begin{center}
   \begin{tabular}{c}
   \includegraphics[height=7cm]{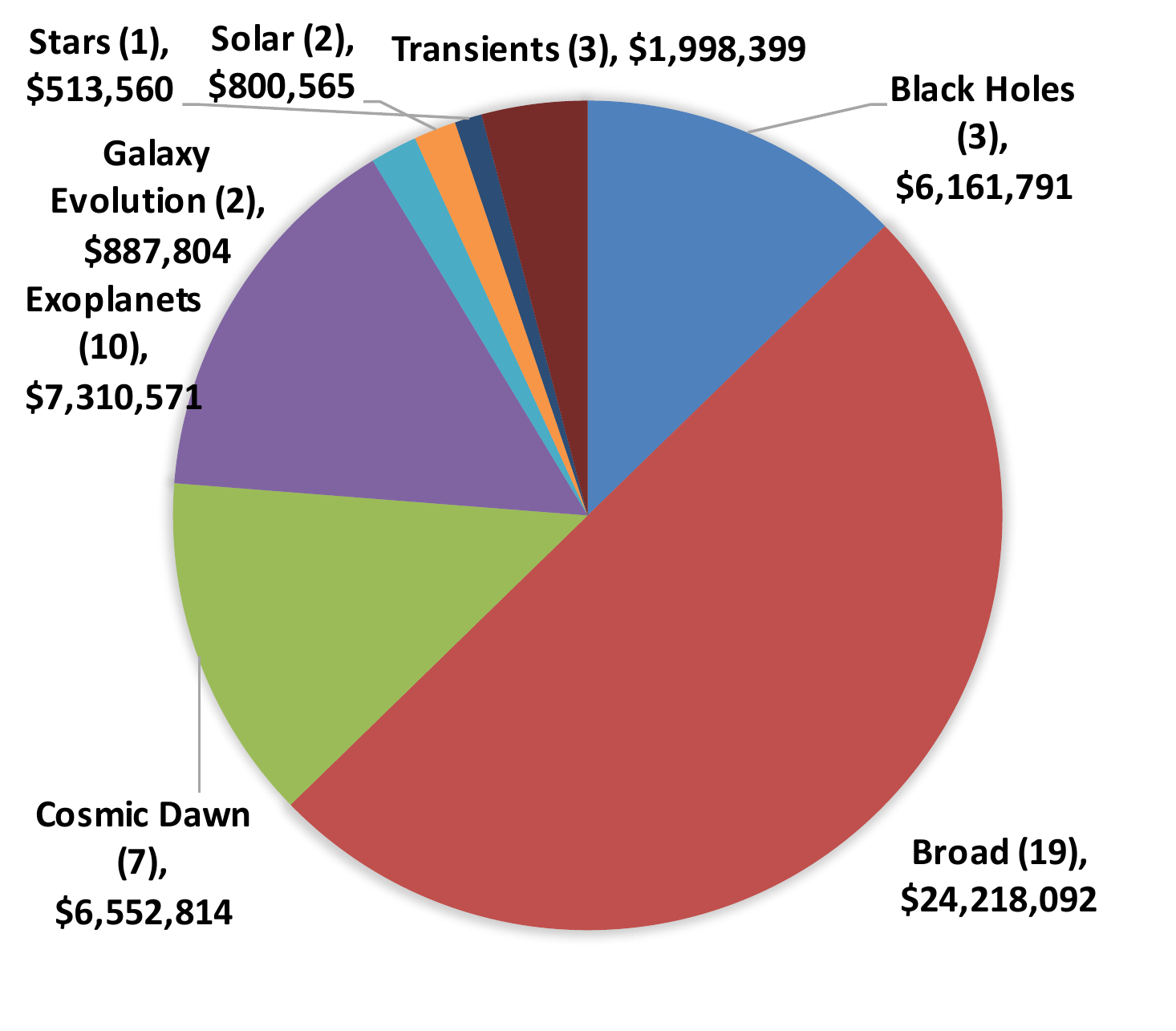}
   \includegraphics[height=7cm]{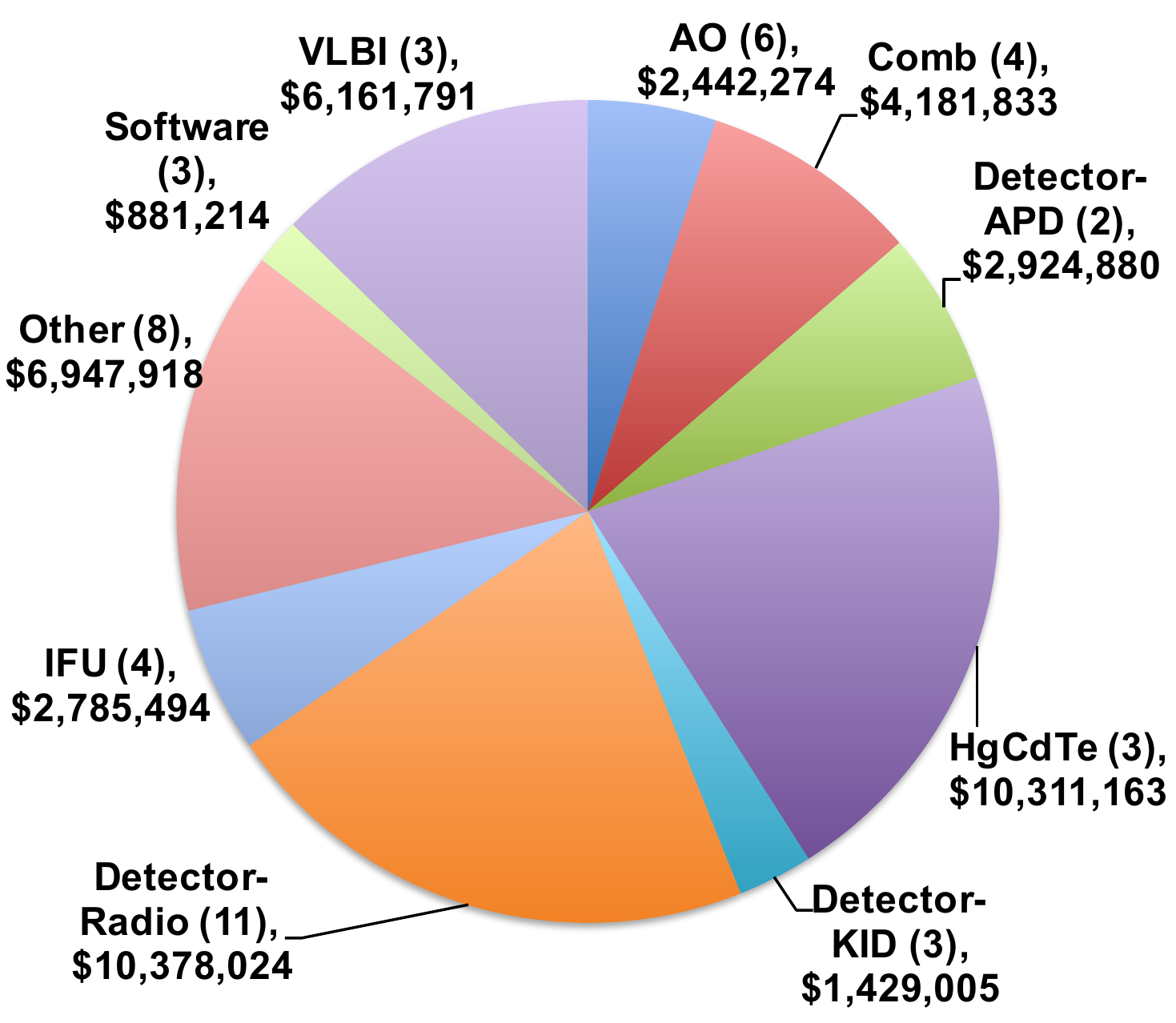}
   \end{tabular}
   \end{center}
   \caption[ActiveAwardsByScienceAndTechnology] 
   { \label{fig: ActiveAwardsByScienceAndTechnology} 
Summary of active ATI awards by science category {\bf (left panel)} and technology type {\bf (right panel)}.  Same format as Figure \ref{fig: ActiveAwardsByWavebandAndTechDevp}.  Awards are distributed across a range of science topics, with the largest segment of awards supporting multiple/broad science goals.  A variety of technologies are supported including Very Long Baseline Interferometry (VLBI), Adaptive Optics (AO), laser frequency combs (Comb), Avalanche Photo-Diode detectors (APDs), HgCdTe IR detectors, Kinetic Inductance Detectors (KIDs), Radio waveband technologies (Radio), Integral Field Units (IFUs), other un-categorized technologies (Other) as well as software.}
   \end{figure} 

A summary of the active ATI portfolio is illustrated in Figures \ref{fig: ActiveAwardsByWavebandAndTechDevp} and \ref{fig: ActiveAwardsByScienceAndTechnology}.  These data were downloaded on October 1, 2016 and therefore do not include awards made in FY 2017 or later. There was no ATI competition in FY 2018.  Awards were selected in the nsf.gov award search by selecting all active awards with program element code 1218, which corresponds to the ATI program. Active awards were categorized by waveband, science objective, technology type and technology development phase based on information in the title and abstract.  Technology development phase is intended to classify the maturity of a technology and it is analogous to Technology Readiness Levels used by NASA\cite{2014SPIE.9143E..13P,Perez2016}.  Awards were classified as ``initial" phase if the proposed research was an initial concept development that would demonstrate a process or technology but would not lead to a full engineering prototype at the end of the award period.  Awards were classified as ``early" phase if they planned to deliver an engineering prototype to demonstrate a technology but would not deliver new science data at the end of the award period.  ``Late" technology development awards would produce new science data at the end of the award period (\eg by installing an instrument at a telescope and using it for science).  

As illustrated in Figure \ref{fig: ActiveAwardsByWavebandAndTechDevp}, the active ATI portfolio is broadly diversified among wavebands and technology development phase; about two-thirds of the program consists of projects to study optical-IR wavebands. About half of the IR waveband expenditures shown in Figure  \ref{fig: ActiveAwardsByWavebandAndTechDevp} are for a single award, 0804651/Hall (\$7M) for development of the HAWAII 4RG detector, discussed below.  Figure \ref{fig: ActiveAwardsByScienceAndTechnology} shows that ATI covers a broad range of science and technology areas.  Exoplanets and cosmic dawn (cosmic dark ages, epoch of reionization and early structure formation) are notable science categories.  Three awards to study black holes (including one award that is cross-listed as an MRI award) are actually in support of one over-arching research program - the Event Horizon Telescope.  Technology categories are similarly broad; the categories most relevant for optical-IR astronomy include adaptive optics (AO; high spatial resolution imaging), laser frequency combs (an enabling technology for high precision radial velocity measurements and exoplanet research), IR detectors (including avalanche photo-diodes, APDs, Mercury Cadmium Telluride (HgCdTe), and Kinetic Inductance Detectors, KIDs) and integral field units (IFUs; spatially resolved spectrographs).

For comparison, the ATI program may be compared with the Planetary Astronomy (PLA) program.  PLA was selected as a control sample because of its comparable budget, number of awards and availability of historical data over a similar range of dates as ATI.  This sample included 445 awards between 1989 - 2016.  Planetary astronomy funds pure science proposals without an emphasis on instrumentation; thus a comparison with ATI may address whether there is a difference in impact for astronomical technology development as opposed to pure science. 

\begin{figure}
   \begin{center}
   \begin{tabular}{c}
  \includegraphics[height=7cm]{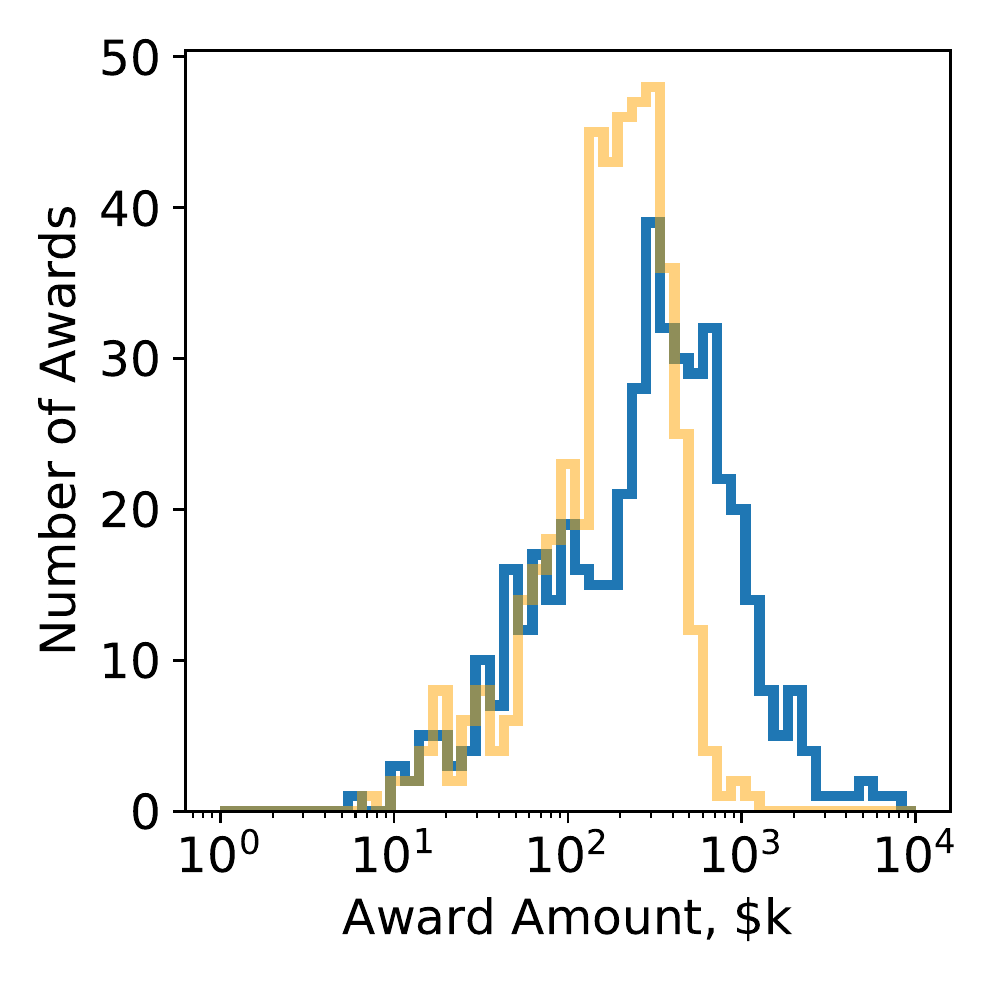}
   \includegraphics[height=7cm]{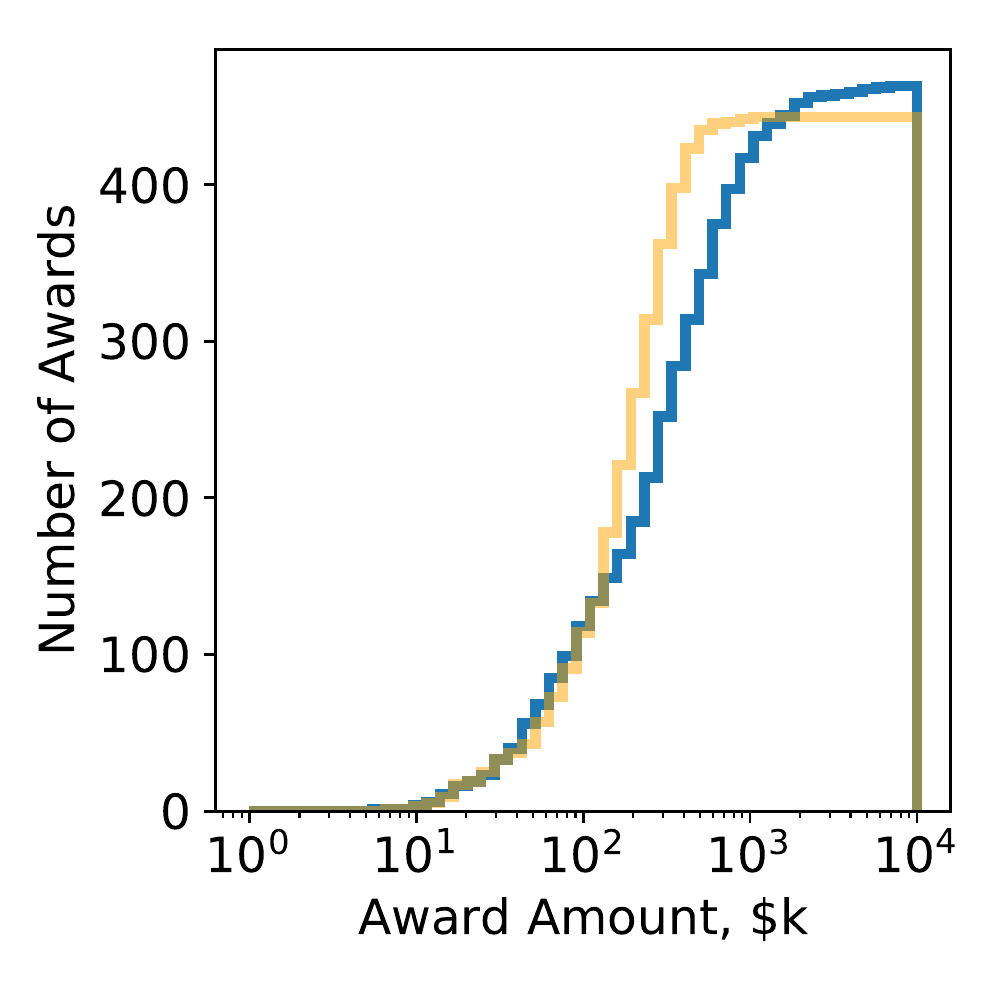}
   \end{tabular}
   \end{center}
   \caption[ATIandPLAAwardDistribution] 
   { \label{fig:ATIandPLAAwardDistribution} {\bf (left panel)}: Distribution of amounts (in thousands of dollars) for all non-zero ATI (blue) and PLA (orange) awards. {\bf (right panel)}: Cumulative distribution of award amounts (in thousands of dollars).}
   \end{figure}

Figure \ref{fig:ATIandPLAAwardDistribution} illustrates distribution of awards amounts in ATI and PLA (in nominal dollars, not inflation adjusted). Over the study period, ATI has awarded a total expenditure of \$222M which is 2.3 times the expenditure for PLA. The median award for ATI (PLA) is \$294K (\$194K). Smaller awards ($<\$50$K) include a variety of categories in addition to research projects, as discussed below. 

\begin{figure}
   \begin{center}
   \begin{tabular}{c}
  \includegraphics[height=7cm]{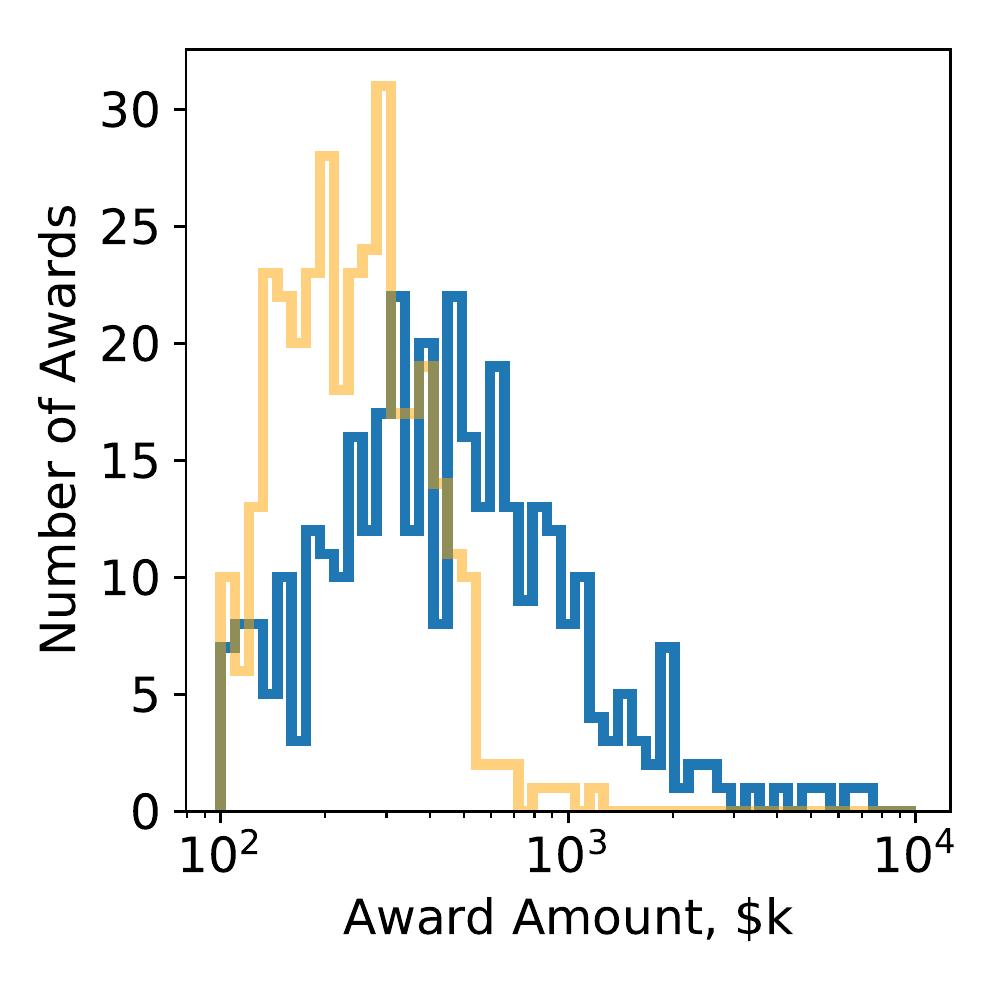}
   \end{tabular}
   \end{center}
   \caption[ATIandPLALargeAwardDistribution] 
   { \label{fig:ATIandPLALargeAwardDistribution} Distribution of large award amounts ( $>$ \$100K; in thousands of dollars) that are typical of new research projects for ATI (blue) and PLA (orange). The mean (median) large ATI award is \$640K (\$421K), and the mean (median) large PLA award is \$270K (\$240K). Thus the typical new ATI research project is about double the cost of a typical PLA project.
}
   \end{figure}

Figure \ref{fig:ATIandPLALargeAwardDistribution} illustrates the subsample of large awards ($>\$100$K). These awards are more typical of new research projects. The mean (median) large ATI award is \$640K (\$421K), and the mean (median) large PLA award is \$270K (\$240K). Thus the typical new ATI research project is about double the amount of the typical PLA project.

As illustrated in Figure \ref{fig:ATIandPLAAwardDistribution}, ATI and PLA have a number of small awards. There are 67 (44) awards in the ATI (PLA) programs that have award amounts less than \$50K. Inspection of titles and abstracts of these awards shows differences in the types of projects that are supported by these small grants. ATI and PLA have comparable percentages of small awards for conferences/workshops (20-25\%) and specific astronomical observations (13\%). However, ATI has a large percentage of small awards for equipment (purchase of specific hardware; 33\%) whereas PLA has hardly any small awards for equipment (2\%). In addition, ATI has a fraction of small awards for miscellaneous categories like teaching and travel (combined 10\%), whereas PLA does not support any of these categories. Instead, the bulk of small PLA awards are for research (solving a specific scientific problem or developing a specialized instrument; 64\%), whereas a smaller fraction of small ATI awards (19\%) are for research. Since 2000, most of the small awards in ATI and PLA are either for conferences or research.

\section{ACKNOWLEDGEMENT AND CITATION ANALYSIS}
\label{LiteratureSection}

\subsection{Automated Literature Search}

An automated search of the astronomical literature was done to asses the direct impact of ATI awards.  Because investigators are required to acknowledge their grants in publications that follow from supported research, it is possible to connect specific research products to particular awards using publicly available literature repositories.  

The sample for the automated search was selected from publicly available data for the ATI program that was available on or about October 4, 2017. Public records for each award consisted of 25 fields of data with information such as NSF award number, managing organization (NSF Division), principal investigator, title, awarded amount to date, start and end dates, program element code(s), award instrument (standard grant, continuing grant, cooperative agreement, or interagency agreement) and abstract as well as other data. The nsf.gov award search page was used to download all public information from awards with program element code 1218 from January 1, 1987 through December 31, 2016. This search yielded results for 582 awards. Many awards receive funding from multiple programs within an NSF Division and sometimes across NSF Divisions. 23 awards were removed from the sample due to their being managed by another Division outside of the Division of Astronomical Sciences (denoted as AST in the NSF Organization field). To select principally ATI awards, awards were removed from the sample that had other program element codes listed first in the case of multiple program element codes or that were obviously not ATI projects based upon the title, abstract or other information (\eg title included MRI designation for Major Research Instrumentation program). The final sample for automated analysis consisted of 496 awards.  

Each award was searched in the astronomical literature using the Astrophysics Data System (ADS).  Code was written in Python to retrieve records from the database for use as input ADS search criteria.  A publicly available interface was used to interact with ADS\cite{AndyCasey}. For each award record, ADS was searched for papers with the following criteria:  (1) the award ID in the full text of the paper, (2) the Principal Investigator (PI) name as an author (3) publication date range specified by the award effective date and 2017.  The search was restricted to peer-reviewed publications.  The decision to exclude conference presentations, even though many instrument papers are published in non-refereed proceedings, was chosen to make the results of the automated search conservative, but more robust. Although some conference proceedings include published manuscripts that allow authors to acknowledge awards, some (notably including American Astronomical Society meetings) include only abstracts that do not typically contain acknowledgments. The number of acknowledging papers as well as their number of citations of each acknowledging paper were saved for each award.  These search results were stored in a PostgreSQL database for analysis. 

\subsection{Award Acknowledgments}

Table \ref{tab:WidelyAcknowledgedAwardsAndCitedPapers} illustrates the most widely acknowledged awards and the most highly cited paper among those acknowledging awards.  The most widely acknowledged ATI award, 0906060/Baranec supported deployment of a micro-machined deformable mirror for the adaptive optics system on the Robo-AO robotic observatory.  This award was acknowledged thirty-one times in the peer-reviewed literature.  The most highly cited paper from this award was cited seventy-nine times\cite{2014ApJ...790L...8B}.  The other two most widely acknowledged awards, 0705139/Ge and 1006676/Mahadevan supported technology for studying exoplanets.  These ATI awards have been broadly aligned with the objectives of the most recent astronomy \& astrophysics decadal review, for which ``seeking nearby, habitable planets" was one of three main science objectives\cite{NWNH2010}. 

%
%
\begin{table}[h]
\caption{Three most widely acknowledged ATI awards.  Column (1) is the NSF award ID.  Column (2) is the Principal Investigator (PI) of the award.  Column (3) is the number of acknowledgements found in the automated search.  Column (4) is the ADS bib code for the most highly cited paper among the acknowledging publications for this award.  Column (5) is the number of citations that the most cited publication has in ADS} 
\label{tab:WidelyAcknowledgedAwardsAndCitedPapers}
\begin{center}       
\begin{tabular}{|c|c|c|c|c|} 
\hline
\rule[-1ex]{0pt}{3.5ex}  {\bf Award ID} & {\bf PI} & {\bf Acknowl.}  &{\bf Most Cited} &{\bf Citations} \\
\rule[-1ex]{0pt}{3.5ex}  {\bf (1)} & {\bf (2)} & {\bf (3)}  &{\bf (4)} &{\bf (5)} \\
\hline
\rule[-1ex]{0pt}{3.5ex} 0906060	&	Baranec, Christoph	&	31	&	2014ApJ...791...35L	&		79\\
\rule[-1ex]{0pt}{3.5ex} 0705139	&	Ge, Jian			&	27	&	2011ApJ...728...32L	&		29\\
\rule[-1ex]{0pt}{3.5ex} 1006676	&	Mahadevan, Suvrath	&	22	&	2014Sci...345..440R	&		68\\
\hline
\end{tabular}
\end{center}
\end{table} 

   \begin{figure}
   \begin{center}
   \begin{tabular}{c}
   \includegraphics[height=7cm]{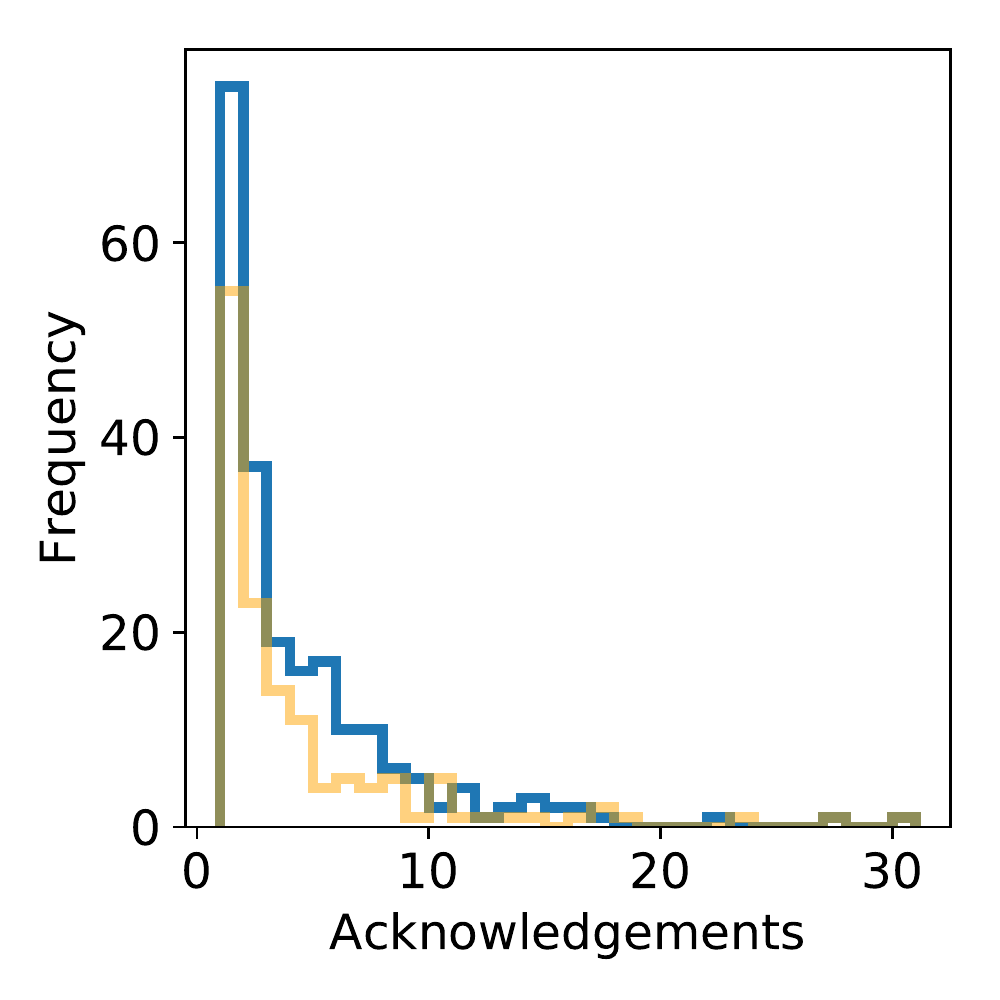}
   \end{tabular}
   \end{center}
   \caption[AwardAcknowledements] 
   { \label{fig:AwardAcknowledements} 
Distribution of acknowledgements of ATI (blue) and PLA (orange) awards. Only awards with at least one acknowledgement are included for clarity. ATI has more acknowledgements than PLA overall, although the difference is not statistically significant (KS test, p-value 0.76).}
   \end{figure} 

Figure \ref{fig:AwardAcknowledements} illustrates the distribution of acknowledgements for the ATI and PLA programs in the peer-reviewed literature. 44\% (216/496) of ATI awards are acknowledged at least once in the peer-reviewed literature, compared with 31\% (138/445) in PLA. Multiple acknowledgements were found with less frequency, and a handful of awards from each program were acknowledged more than twenty times. Figure \ref{fig:AwardAcknowledements} demonstrates that ATI and PLA awards are acknowledged with approximately the same frequency in the peer-reviewed astronomical literature despite the fact that ATI awards emphasize technology development, whereas PLA awards are for ``pure science." 

\begin{figure}
   \begin{center}
   \begin{tabular}{c}
   \includegraphics[height=7cm]{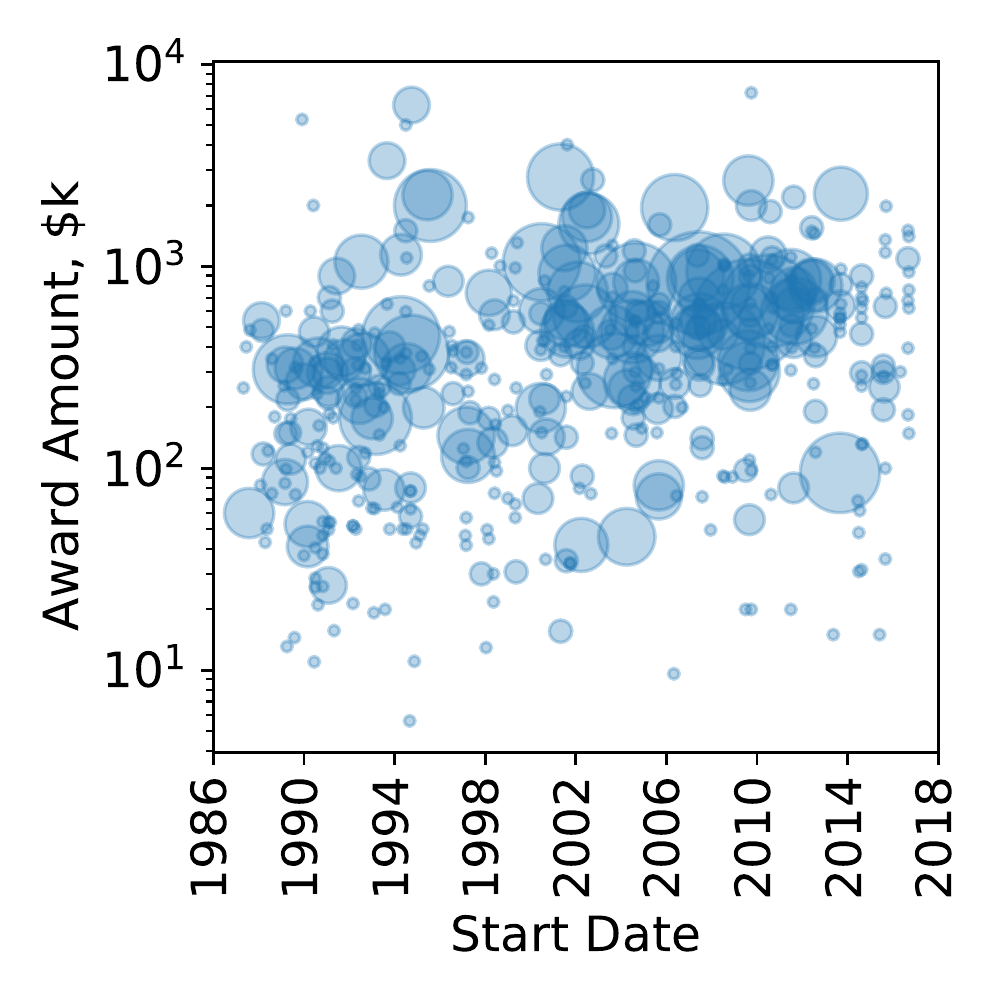}
   \includegraphics[height=7cm]{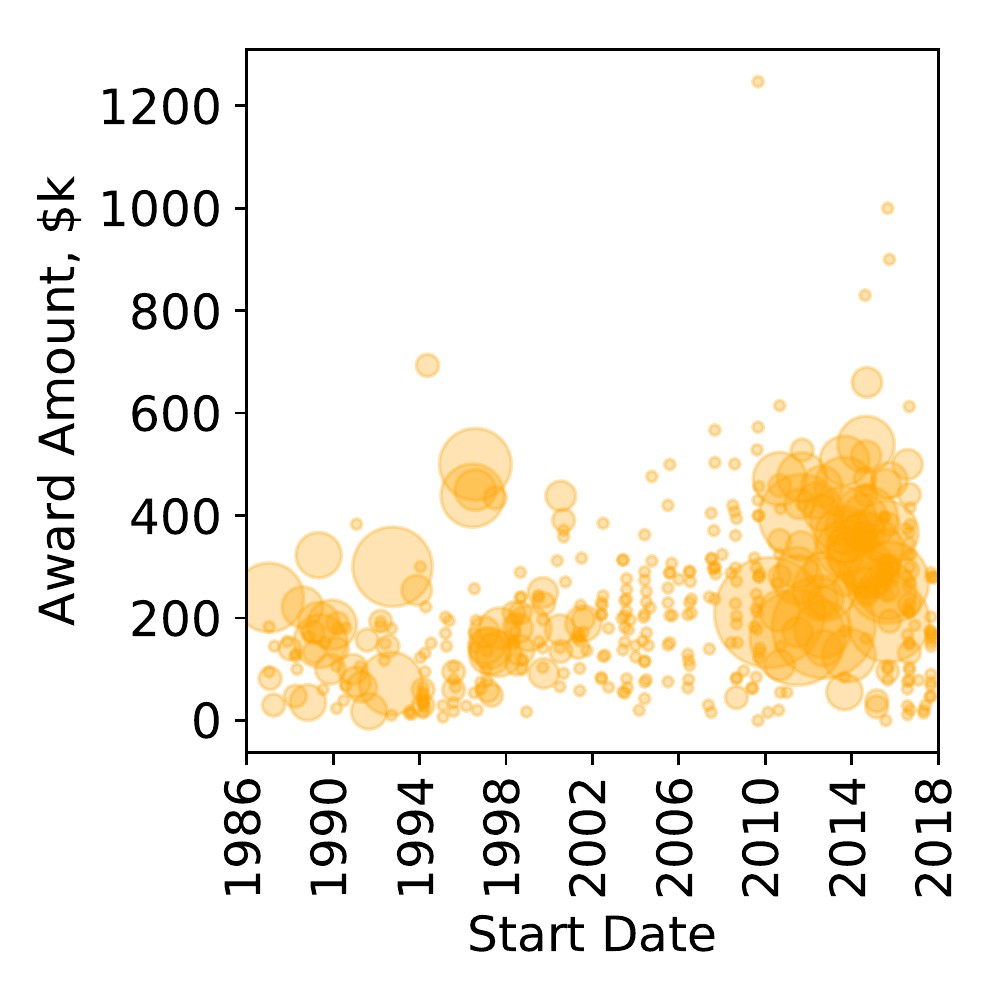}
   \end{tabular}
   \end{center}
   \caption[AwardAcknowledements] 
   { \label{fig:AwardAmountDateAcknowledgement} 
Award amount (in thousands of dollars) vs start date for ATI (left panel; blue symbols) and PLA (right panel; orange symbols). Symbol size corresponds to the number of ADS acknowledgements}
   \end{figure}

Figure \ref{fig:AwardAmountDateAcknowledgement} illustrates award amounts over the study period for both programs, where the symbol size corresponds to the number of acknowledgements of each award. This figure shows that widely acknowledged awards (large symbols in the figures) have occurred in each program throughout the past thirty years, although there was a notable deficit of acknowledgements in PLA in the decade 2000-2010. Figure \ref{fig:AwardAmountDateAcknowledgement} also illustrates that wide acknowledgement can be found among both large and small awards. There is no strong trend between award amount and acknowledgement.

\subsection{Literature Citations}

Table \ref{tab:CitationsAndImpact} shows the five most highly cited papers that acknowledge their awards from the last thirty years of the ATI and PLA programs.  The most highly cited ATI paper presents first results of the Degree Angular Scale Interferometer (DASI) measurement of the cosmic microwave background angular power spectrum\cite{2002ApJ...568...38H}.  The receivers, the dewars, the local oscillator distribution scheme, and the correlator and were all designed and developed with support from  9413935/Readhead in building the Cosmic Background Imager (CBI) of which DASI was a copy.  

%
%
\begin{table}[h]
\caption{The most highly cited papers that acknowledge ATI and PLA awards.  Columns (1,2) indicate the NSF award ID and PI last name respectively.  Column (3) is the ADS identifier for the paper.  Column (4) is the year of publication of the paper.  Column (5) is the number of citations.  Column (6) is the median number of citations for the year of publication.  Column (7) is the impact, defined as citations / median.  * indicates an unreliable value as discussed in the text.} 
\label{tab:CitationsAndImpact}
\begin{center}       
\begin{tabular}{|c|c|c|c|c|c|c|} 
\hline
\rule[-1ex]{0pt}{3.5ex}	{\bf Award}	&	{\bf PI}	&	{\bf Bibcode}	&	{\bf Year}	&	{\bf Citations}	&	{\bf Median}	&	{\bf Impact}	\\
\rule[-1ex]{0pt}{3.5ex}	{\bf (1)}	&	{\bf (2)}	&	{\bf (3)}	&	{\bf (4)}	&	{\bf (5)}	&	{\bf (6)}	&	{\bf (7)}	\\
\rule[-1ex]{0pt}{3.5ex}		&		&	{\bf ATI}	&		&		&		&		\\
\rule[-1ex]{0pt}{3.5ex}	9413935	&	Readhead	&	2002ApJ...568...38H	&	2002	&	769	&	36	&	21.4	\\
\rule[-1ex]{0pt}{3.5ex}	0096913	&	Carlstrom	&	2002ARA\&A..40..643C	&	2002	&	561	&	36	&	15.6	\\
\rule[-1ex]{0pt}{3.5ex}	8822465	&	McCarthy	&	1991ApJS...77..417K	&	1991	&	495	&	25	&	19.8	\\
\rule[-1ex]{0pt}{3.5ex}	0904607	&	Townsend	&	2013ApJS..208....4P	&	2013	&	489	&	9*	&	54.3*	\\
\rule[-1ex]{0pt}{3.5ex}	9203336	&	McCarthy	&	1993AJ....106..773H	&	1993	&	420	&	27	&	15.6	\\
\rule[-1ex]{0pt}{3.5ex}	 	&	 	&	 	&	 	&	 	&	 	&	 	\\
\hline
\rule[-1ex]{0pt}{3.5ex}		&		&	{\bf PLA}	&		&		&		&		\\
\rule[-1ex]{0pt}{3.5ex}	9120599	&	Begelman	&	1994ApJ...421..153S	&	1994	&	860	&	27	&	31.9	\\
\rule[-1ex]{0pt}{3.5ex}	8857365	&	Wisdom	&	1991AJ....102.1528W	&	1991	&	675	&	25	&	27.0	\\
\rule[-1ex]{0pt}{3.5ex}	9530590	&	Heiles	&	2003ApJ...586.1067H	&	2003	&	341	&	28	&	12.2	\\
\rule[-1ex]{0pt}{3.5ex}	9973057	&	Tedesco	&	2002AJ....123.1056T	&	2002	&	310	&	36	&	8.6	\\
\rule[-1ex]{0pt}{3.5ex}	9714275	&	Lin	&	2001ApJ...548..466B	&	2001	&	249	&	31	&	8.0	\\
\rule[-1ex]{0pt}{3.5ex}	 	&	 	&	 	&	 	&	 	&	 	&	 	\\
\hline
\end{tabular}
\end{center}
\end{table} 

In addition to the number of citations and the year of publication for each paper, Table \ref{tab:CitationsAndImpact} also shows the median number of citations for that year of a paper published in {\it The Astronomical Journal}.  This median value was obtained from a search of ADS for all papers in {\it The Astronomical Journal} for the year in question.  Notably, median citations in this journal are in the 20-30 range throughout the 1990's and 2000's, but they decline significantly after 2009.  Therefore, after 2009 median citations in this journal should not be taken as representative of the entire field (applies to one paper in the table, entry marked with an asterisk).  

An impact factor is defined as the number of citations divided by the median number of citations for that year, and it is tabulated in the last column of Table \ref{tab:CitationsAndImpact}.  The tabulated impact factor after 2009 is considered unreliable due to uncertainty in the median.  The impact factor normalizes for an age effect, whereby older papers have more citations simply because they are older.  The distribution of such normalized citation counts has been found to follow the same log-normal distribution across a wide range of scientific disciplines\cite{Fortunato:2018ih,2008PNAS..10517268R}.  The papers shown here are in the high-impact tail of this distribution.  

We compared the impact distribution of ATI papers both to astronomy as a whole and to PLA papers specifically. To establish the impact distribution of the entire astronomy literature, publications from six journals were used including {\it Astronomy \& Astrophysics}, {\it The Astrophysical Journal}, {\it The Astronomical Journal}, {\it The Astrophysical Journal Supplement}, {\it Monthly Notices of the Royal Astronomical Society}, and {\it Publications of the Astronomical Society of the Pacific}. All 157,039 articles from 1988-2015 from these six publications were used to determine the astronomy literature citation distribution. Citations up to 2017 were used. 

To get normalized distributions, the counts of the raw, unbinned distributions were divided by the number of papers published that year. Normalizaton means that the sum of all $y$ values will be one. The distributions were first normalized in order to address different productivity levels in different years. Normalization achieves that it does not matter if one distribution (for some year) had 1000 articles and another 10,000. If the fraction of articles with $x$ citations is the same, these distributions are the same. For each year, the the mean and the median of the number of citations were computed. This is to take into account that older publications had more time to accumulate citations. 

\begin{figure}
   \begin{center}
   \begin{tabular}{c}
   \includegraphics[height=7cm]{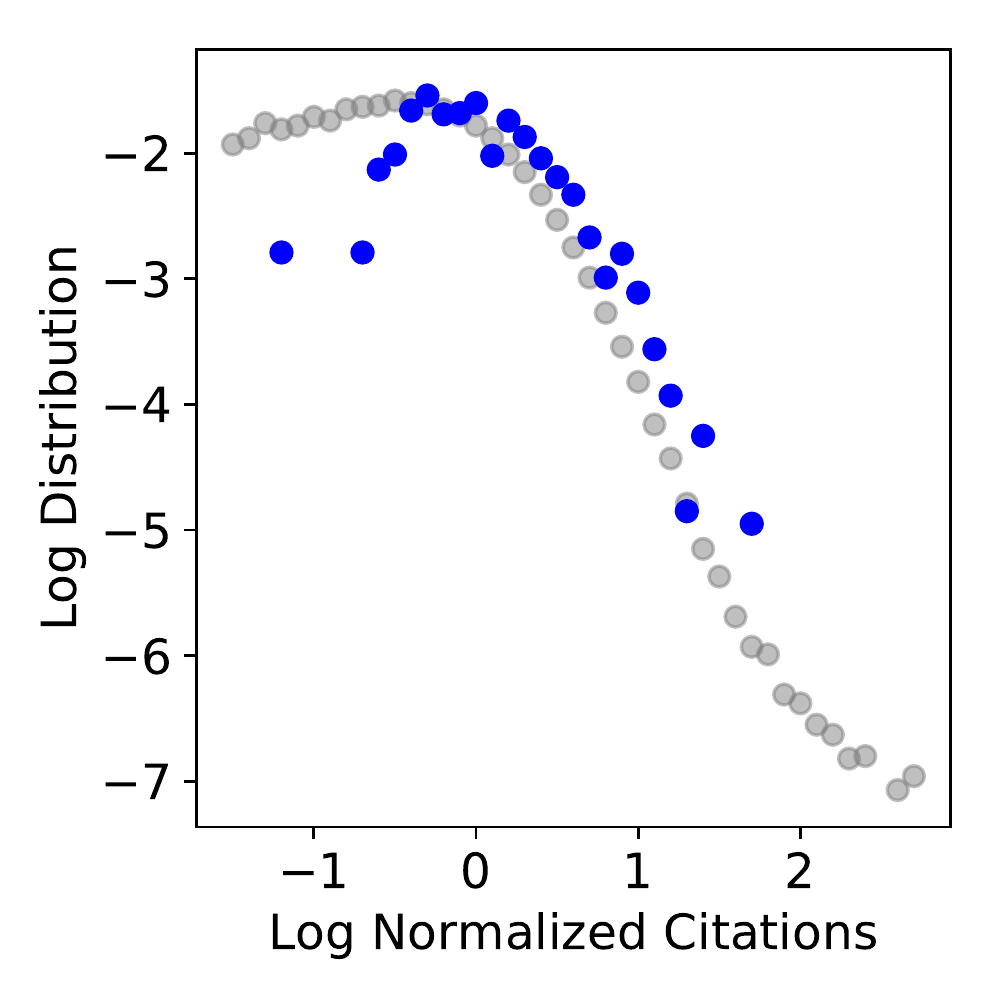}
   \includegraphics[height=7cm]{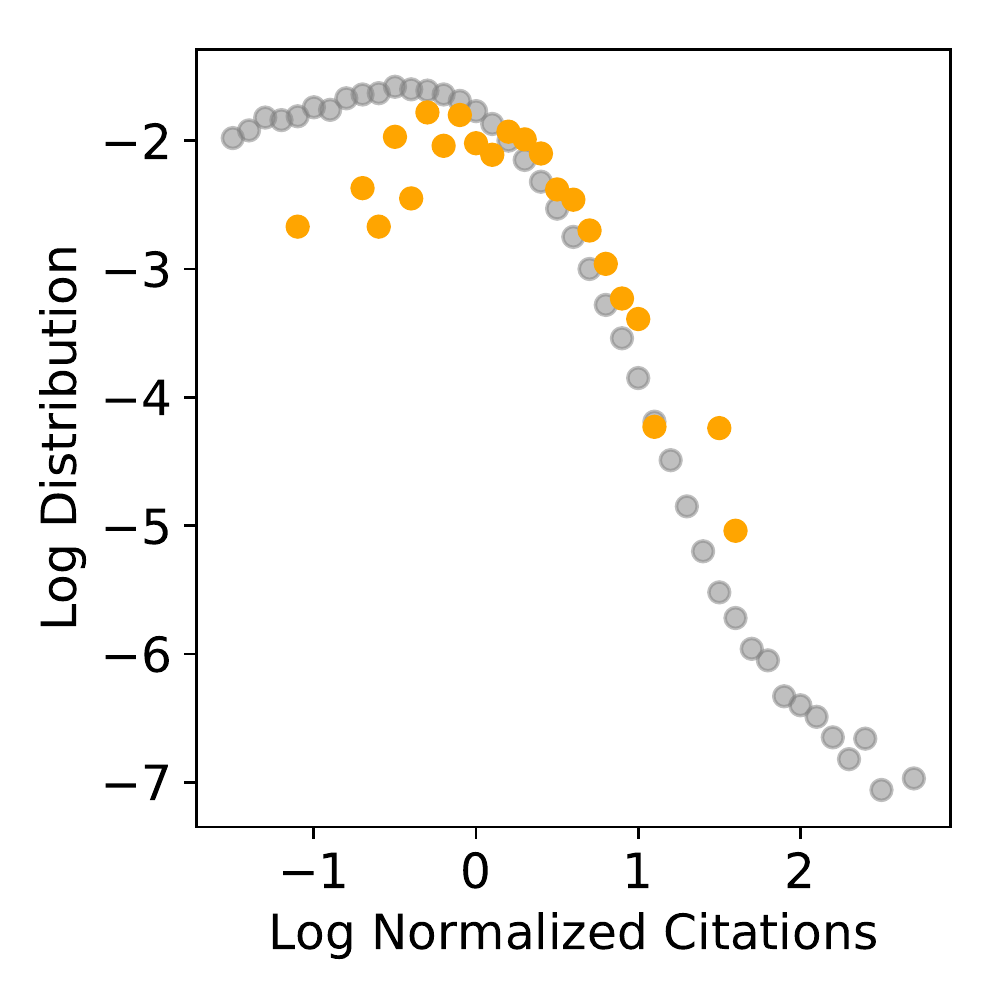}
   \end{tabular}
   \end{center}
   \caption[CitationDistribution] 
   { \label{fig:CitationDistribution} 
Distribution of impact (median normalized citations) of papers from ATI (left panel; blue circles) and PLA (right panel; orange circles) programs compared to the aggregate astronomy literature (gray circles). Impact has been found to track a universal distribution across a wide range of scientific disciplines. ATI and PLA have very similar impact distributions despite ATI being focused on technology and instrumentation development in addition to pure science. ATI and PLA both have fewer low impact papers and somewhat more high impact papers than the universal distribution.}
   \end{figure}

The distributions of citations for each year were logarithmically binned using 0.1 decade bins.  Then every individual citation distribution was normalized by the median to get the universal distribution. This basically means that the $x$ axis was shifted by log(median), to take into account that older publications had more time to accumulate citations. Finally these citations were averaged for all the years in the dataset. Averaging is done by summing up the values of distributions (at different logarithmic bins) and dividing by the number of distributions (\ie the number of different years). Figure \ref{fig:CitationDistribution} plots the logarithm of the probability distribution of citation vs the logarithm of median-normalized citations. The citation distribution for astronomy is illustrated by the gray symbols in this figure.

Figure \ref{fig:CitationDistribution} illustrates the citation distributions for papers that acknowledge an ATI award (left panel; blue symbols) and a PLA award (right panel; orange symbols). These distributions are qualitatively similar to the general astronomy distribution (gray symbols) at medium and high impact values (\ie the curves have a similar functional form for values of logarithm of normalized citations greater than zero).  However, there are more high impact papers among ATI and PLA than in the general astronomical literature, \ie the blue symbols lie above the gray symbols toward the right half of the plots. Furthermore, there are comparably fewer low impact papers (logarithm of normalized citations less than zero) among the ATI and PLA samples than in the general astronomical literature, \ie the blue symbols are below the gray symbols toward the left side of the plots. Thus ATI and PLA-supported research outperforms the general astronomical literature in terms of literature citations. We interpret this effect as a selection bias resulting from the peer review process: proposals that are awarded by NSF have been selected through competition.

As illustrated in Figure \ref{fig:CitationDistribution}, the impact distributions of ATI and PLA are essentially the same. Technology development and instrumentation papers are no less impactful in the scientific literature than pure science without a technology or instrumentation component. This surprising conclusion countermands the naive intuition that might have supposed that instrument builders devote less time and effort to writing papers because they spend more time in the laboratory. In fact, papers from technology and instrumentation programs are just as impactful as pure science papers.

%

\subsection{Sources of Uncertainty}

Several factors may affect the accuracy of literature acknowledgements and resulting paper citations as well as their interpretation.  Papers may acknowledge multiple grants, and it can be impossible to determine solely from published literature how much or how critical was the contribution of any one grant to a larger project. Indeed, in rare cases, grants are acknowledged in papers on subjects that are not obviously aligned with the original grant. Research is a creative process that can lead to unexpected directions; furthermore, progress in one field may enable serendipitous advances in other fields.

Acknowledgements are also subject to certain types of error. A false positive acknowledgement (type I error) would be an instance of the automated search reporting a paper to acknowledge an award when in fact no such acknowledgement exists.  No such errors were found among manual checks of several dozen papers selected at random from the search results.  

A false negative (type II error) would be an instance of a paper that was in fact derived from grant-supported research not being found by the automated search.  There are several possible explanations for such false negatives. A product of research may be published outside of the ADS accessible repository; these publications may be books or journals that are not included within ADS.  Alternatively there may be a missing or incomplete acknowledgement of an NSF award; such delinquent acknowledgements are the most common source of false negatives. They may arise if the award number is not included in the acknowledgement, or if the acknowledged award number is incorrect, or if the authors neglect to include any acknowledgement. Due to all false negatives, acknowledgements from the automated search are believed to underestimate the true numbers of publications by approximately a factor of two. 

\section{DISCUSSION:  ATI THROUGH THE YEARS}
\label{DiscussionSection}

The history of the ATI program provides a window into the extraordinary scientific and technological advance of the past thirty years. While it is impossible to review the outcome of each and every ATI award, this section reviews some of the major themes and trends. This section places them within their larger historical and scientific context. From such a longitudinal perspective emerges one of the main conclusions of this paper: the true impact of technology development for astronomy can only be appreciated by taking a long view that extends beyond the lifetime of any individual grant. Technology development in astronomy unfolds over decades.

\subsection{Charge Coupled Device Cameras}

Among the most the most transformative devices for astronomy in the past half century is the Charge Coupled Device (CCD) camera. CCDs were invented in 1969 by Willard S. Boyle and George E. Smith at Bell Laboratories, for which they awarded the 2009 Nobel Prize in Physics.\cite{Boyle1970} A CCD is a metal oxide semiconductor device that stores charge in well-defined regions of the semiconductor and moves it according to applied voltages. The first CCD imaging device consisted of a 1D scanner that had 8 pixels.\cite{2010RvMP...82.2307S} 

In the 1980's, CCDs brought about a revolution in imaging for astronomy.\cite{1986JOSAA...3.2131T} By the early 1990's, several companies had produced CCDs that were suitable for scientific applications; they featured low noise, negligible dark current and high charge-transfer efficiency. However, such devices had low overall quantum efficiency, which was a disadvantage for astronomical applications. Several ATI awards to the University of Arizona (9121801/Lesser, 9500290/Lesser, 9618760/Lesser) supported  post-processing of commercial detectors to optimize their performance for astronomy. Devices were thinned for backside illumination, backside charging and antireflection coating. Such techniques quickly matured and became routinely applied to detectors of various formats and manufacturers. A further ATI award (9876630/Lesser) explored the use of a competing technology, Complementary Metal Oxide Semiconductor (CMOS) detectors, for astronomical use. 

Other ATI awards in the 1980's and 1990's were notable for disseminating CCD cameras to astronomical observatories for research and teaching.  Maunakea, Lick and Lowell as well as many smaller observatories received CCD cameras through ATI awards in this time period. CCD sensors are now the most common sensors in use for visible light astronomy\cite{2015PASP..127.1097L}. 

The camera for the Vera C. Rubin Observatory (formerly Large Synoptic Survey Telescope, LSST) includes CCD sensors with combined 3.2 GigaPixels\cite{2010SPIE.7735E..0JK}; it represents an improvement in pixel count by more than eight orders of magnitude compared to the original CCD imager. LSST focal plane development was initiated in 2002 with an ATI award (0243144/Tyson) which supported a trade study between CCD and CMOS detectors for LSST.  This award was the first NSF investment in what ultimately became the highest-priority for ground-based astronomy, according to the 2010 decadal review.  The ATI award came at a critical phase in the growth of the project; without this essential seed funding, the focal plane may not have ever been developed.\cite{TysonPrivateCommunication} Subsequent development of the LSST camera was supported by the Department of Energy.

\subsection{Large Optical Telescope Mirrors}

The optical telescope, the most iconic symbol of astronomy, has undergone a revolution in design since the 1970's. New approaches to building the primary mirror have enabled transformative growth in the size and hence the light gathering power of the world's largest telescopes. Three techniques for have been particularly successful for making big glass.\cite{McCray2004} A bend and polish method of fabricating segments that fit together to form a large mirror was pioneered by Jerry E. Nelson at the Lawrence Berkeley National Laboratories in the 1970's. This method was ultimately used for the twin 10-m telescopes of the W.~M.~Keck Observatory, and is planned for the Thirty Meter Telescope. Thin meniscus mirrors whose lack of stiffness is dynamically compensated to maintain their optical figures were developed by commercial interests including Corning, which has a history of making mirrors for spy satellites. Such mirrors were employed on the Gemini and Subaru telescopes and the European Very Large Telescope.

\begin{figure}
   \begin{center}
   \begin{tabular}{c}
   \includegraphics[width=3.5in]{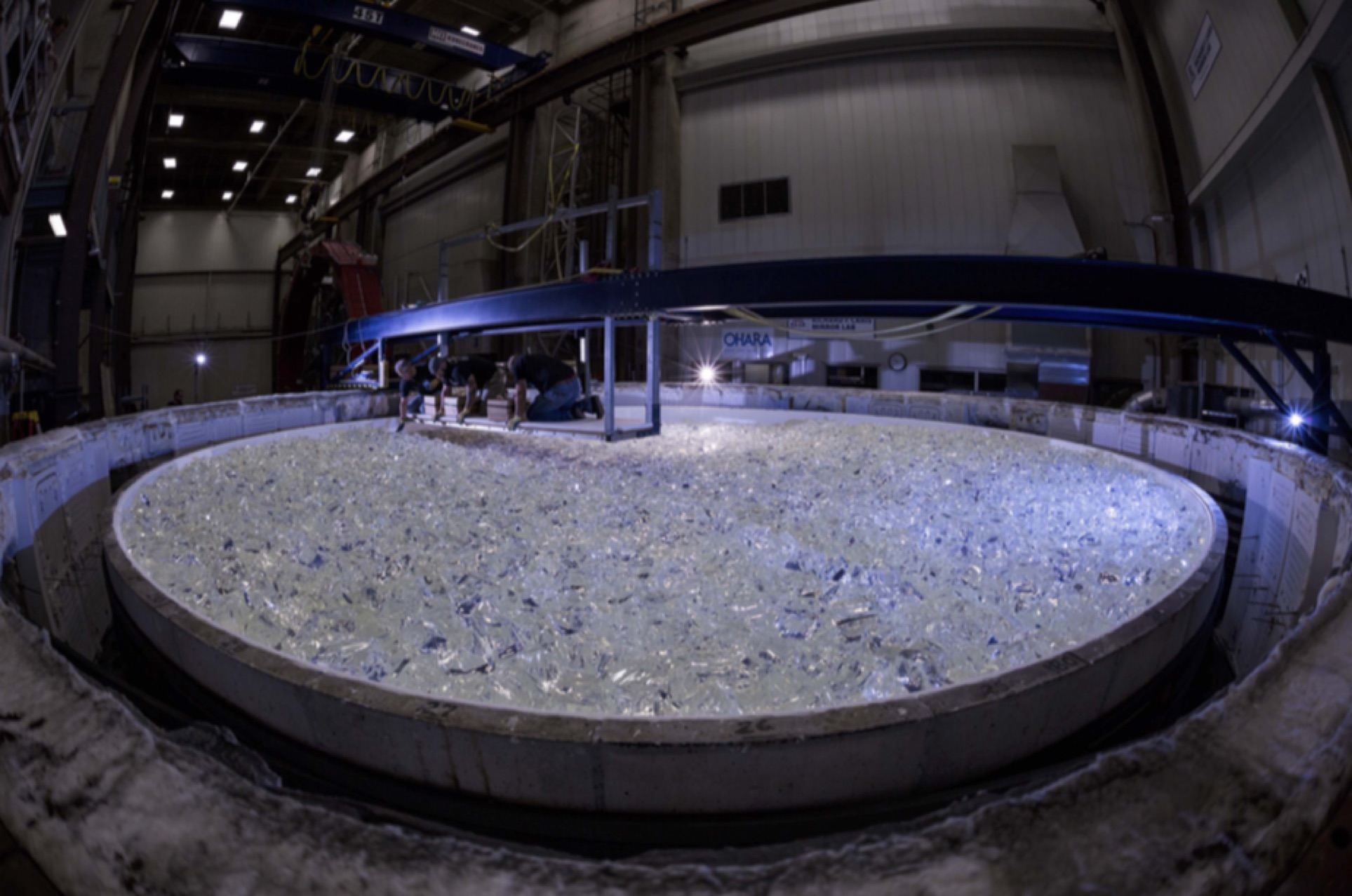}
   \includegraphics[width=3.5in]{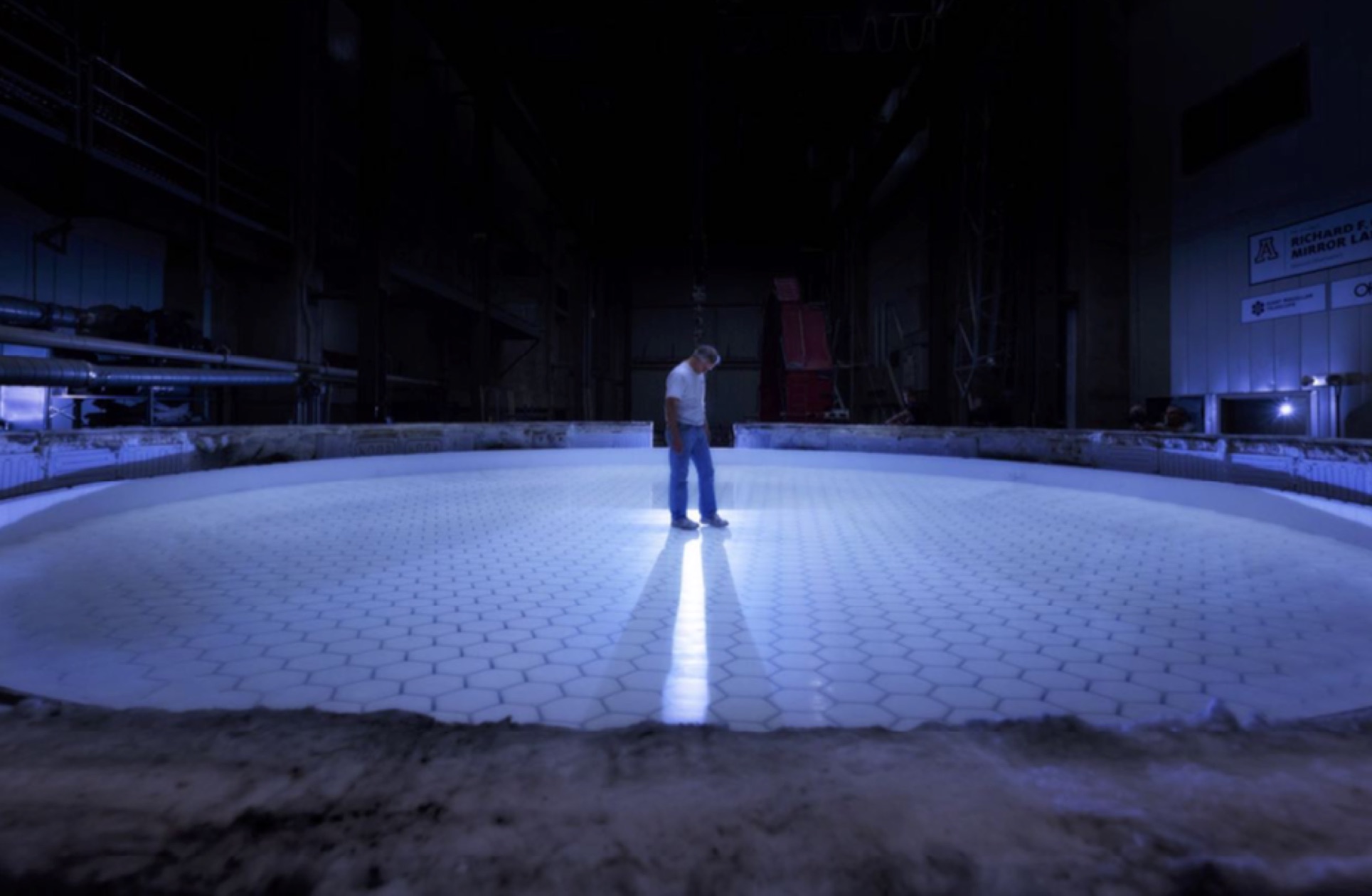}
   \end{tabular}
   \end{center}
   \caption[MirrorLab] 
   { \label{fig:MirrorLab} 
Large optical mirror fabrication using spin-cast, honeycomb mirrors was developed with early support from ATI {\bf (left panel)} Loading 17.48 tons of E6 borosilicate glass {\bf (right panel)}. Inspecting the surface of the mirror after the annealing process and the furnace has been removed. Image credit: Damien Jemison, GMTO, Richard F. Caris Mirror Lab, The University of Arizona}
   \end{figure}

An alternative technique, spin cast mirrors, was pioneered by J.~Roger P.~Angel at the University of Arizona and it was supported in part by NSF/ATI. This technique grew out of backyard experiments with a home-made kiln. The successful spin casting method was ultimately pursued in a dedicated laboratory at the University of Arizona, see Figure~\ref{fig:MirrorLab}. ATI award 8911701/Angel provided \$5M of early support for this lab; other funders included the Air Force Office of Scientific Research and private institutional funds. This lab provided primary mirrors for the MMT, the Large Binocular Telescope (LBT) and the New Solar Telescope for the Big Bear Solar Observatory and is being used to make mirrors for the 30-m class Giant Magellan Telescope (GMT) that is currently under construction as well as others. 

ATI support also provided early seed funding for an early precursor to GMT. 0138347/Angel provided \$1.9M to support early conceptual design and development including mechanical and optical components.\cite{2003SPIE.4840..183A,2004SPIE.5494...62M} The ATI program is conceived as an incubator for potentially transformative technology concepts for ground based astronomy - relatively small awards to seed ideas that may one day transform astronomy. In these cases, ATI awards have indeed seeded initiatives that have subsequently grown to be large astronomical facilities. 

%
%

\subsection{Adaptive Optics}

Ever since telescopes turned toward the sky, astronomers realized that atmospheric turbulence degrades their images.\cite{1999aoa..book.....R} Constrained by the fickle air above, even the most sophisticated ground-based observatory is hobbled to resolution that is not much better than a backyard telescope. This problem seemed insurmountable until 1953 when Horace W. Babcock outlined a solution that came to be known as adaptive optics.\cite{1953PASP...65..229B} The broad outlines were laid bare in his seminal paper: a wavefront sensor detects time-varying optical deformations by locking onto a nearby, bright reference star. Distortions are corrected by an equal and oppositely deforming mirror that operates in closed loop with the wavefront sensor. The method can work for science targets that are fortuitous enough to be nearby a bright reference guide star; this main limitation is effectively one of sky coverage. 

The modern application of this technique is known as Natural Guide Star Adaptive Optics (NGS-AO) to differentiate it from variants that spawned in the succeeding decades. These variants include Laser Guide Star Adaptive Optics (LGS-AO), which uses steerable laser beacon(s) to establish a reference for wavefront correction near the science target. Ground Layer Adaptive Optics (GLAO) offers partially improved seeing over a wide field of view by compensating only for turbulence within $\sim$100 m above the telescope. Multi-Conjugate Adaptive Optics (MCAO) uses multiple wavefront sensors and deformable mirrors to correct turbulence from multiple levels in the atmosphere. Multi-Object Adaptive Optics (MOAO) corrects images of multiple science targets simultaneously. Extreme Adaptive Optics (ExAO) attains high order correction over small fields of view near a bright reference star to obtain extremely high contrast, especially for exoplanet imaging.

The first demonstrations of AO were done in the 1970's.\cite{1978IEEEP..66..651H} Early research was supported by the Air Force and the Defense Advanced Research Projects Agency (DARPA) for space surveillance applications.\cite{1998aoat.book.....H}  However, the first published application of AO to astronomy was the ESO supported system called COME-ON, which was successfully tested in 1989 at Observatoire de Haute Provence in France;\cite{1989Msngr..58....1M,1990A&A...230L..29R} it was later upgraded and moved to the 3.6-m ESO observatory at La Silla, Chile.\cite{1994SPIE.2201.1088R} 

In the 1980's early U.S. development of AO for astronomy continued at NOAO and University of Hawaii. An innovative approach to wavefront sensing and control, the curvature sensor and bimorph mirror, were developed during this time.\cite{1988ApOpt..27.1223R} NSF/ATI award 9319004/Roddier partially supported the deployment of this technology at the Canada France Hawaii (CFHT) telescope at Maunakea, and its use to study circumstellar environments and protoplanetary disks.\cite{1991SPIE.1542..248R,1996ApJ...463..326R} This curvature sensing approach was later adopted by the 8-m class Subaru telescope.\cite{2004PASJ...56..225T} 

\subsubsection{Laser Guide Stars}

Beginning in 1981, Department of Defense (DoD) classified research aimed to overcome the sky coverage limitation of AO. The concept of using a laser as a steerable reference beacon in the sky was first conceived as part of this work. From the vantage of the telescope, an apparent Laser Guide Star (LGS) is made  by an outgoing laser beam pointing at the science target; backscattered laser light from the atmosphere is imaged and used as a reference for wavefront correction. Lasers that induce Rayleigh backscattering or resonant scattering from sodium atoms (at wavelength 0.589~$\mu$m) can be made to work.\cite{1994JOSAA..11..263H} Rayleigh backscattering occurs in the lower atmosphere ($\sim$10's km) whereas resonant scattering from the sodium layer occurs much higher in the atmosphere ($\sim$ 90-100 km). The higher scattering altitude means that sodium lasers suffer less focal anisoplanatism (cone effect). Focal anisoplanatism is a source of systematic wavefront correction error that arises due to the difference in the (cone shaped) volume of atmosphere traversed by the wavefront from the foreground LGS compared to the (cylindrical) volume of atmosphere traversed by light from the background science target.

LGS-AO systems were implemented at DoD sites including the Starfire Optical Range in New Mexico and the Air Force Maui Optical Site in Hawaii.\cite{1991Natur.353..144F}  NSF played a role in declassifying this research and transferring it to the astronomical community. Whether this research should be declassified was a subject of considerable deliberation within the government. Advocates of declassification included key scientists involved in the research as well as members of the Jasons scientific advisory group who had been involved in its inception. NSF/ATI Program Director Wayne Van Citters, who had been briefed on the LGS research, argued that declassification would save years of research and millions of dollars that would not have to be spent by astronomers reinventing work that had been done by DoD scientists.\cite{Duffner2009} In 1985, LGS had been independently proposed in the public literature by Foy and Labeyrie.\cite{1985A&A...152L..29F} AO would be necessary for coming generations of large optical telescopes, and it was anticipated that further innovations by astronomers would improve the technology that could ultimately benefit even the originating DoD programs. Declassification was ultimately approved in 1991.

Upon declassification of this research, NSF/ATI played a key role in disseminating the technology for use in astronomy. Awards that facilitated application of DoD facilities for astronomy included an interagency agreement with the Air Force (9500168/Fugate) as well as twenty one grants to support astronomical observations at AEOS under a special NSF solicitation. ATI also supported numerous conferences and workshops for adaptive optics over the years.

ATI supported the development of laser guide stars and the deployment of AO instrumentation at Maunakea, Mt.~Wilson and Apache Point observatories and the Multiple Mirror Telescope (MMT) beginning in the 1990's. Other contemporary early LGS-AO systems were developed for the 3-m Shane telescope at Lick Observatory\cite{1997Sci...277.1649M} developed in conjunction with Lawrence Livermore National Laboratory and the 3.5-m telescope at Calar Alto Observatory\cite{2000ExA....10..103D} in Spain. 

Using the MMT, ATI award 9203336/McCarthy supported the first demonstration of LGS-AO for astronomy as part of a research program through the University of Arizona.\cite{1995ApJ...439..455L} Analysis of corrected images were consistent with theoretical predictions and this gave confidence that LGS-AO systems would indeed have tremendous value for astronomy. This research program went on to demonstrate the first correction of an extended image with LGS-AO. Supported by additional ATI awards (0138347/Angel and 0505369/Hart), in 2005 they demonstrated the first use of multiple laser guide stars,\cite{2005ApJ...634..679L} and in 2009 they demonstrated successful implementation of GLAO.\cite{2009ApJ...693.1814B,2010Natur.466..727H} This research was a key milestone in the development of MCAO, discussed below.

Focal anisoplanatism becomes particularly restrictive for larger apertures, and it requires the largest aperture telescopes to use sodium LGS as opposed to Rayleigh LGS; this technical requirement motivated much research into making lasers that operate at the sodium resonance of 0.589~$\mu$m. Edward Kibblewhite received four ATI awards between 1993 and 2008 (9256606, 9421406, 9731169, 0837646); this research program developed the sum frequency laser guide star (with Lincoln Labs/Tom Jayes).  Although not adopted in the current lasers (produced by Toptica Photonics), this program was a key evolution to the current generation of laser guide star systems and included developing a physical understanding of the impact of various laser formats; it was also the laser type used during early science with AO systems on Gemini, Keck and Subaru.\cite{ChunPrivateCommunication}  

Major observatories deployed LGS-AO systems beginning in the 1990's, notably including the W.M. Keck Observatory\cite{1993ARA&A..31...13B}, which first used LGS-AO in 2003. Gemini Observatory, the Very Large Telescope (VLT), Subaru Telescope, Palomar Observatory, the Large Binocular Telescope, Southern Astrophysical Research Telescope (SOAR), William Herschel Telescope and others developed LGS-AO systems. 

\subsubsection{Deformable Mirrors}

In most AO systems, wavefront correction is achieved with a deformable mirror. As AO technology matured in the 1990's and 2000's, piezoelectric actuated deformable mirrors with $\sim$100 actuators were the most common implementation. However, the development of Micro-ElectroMechanical Systems (MEMS) offered the potential of semiconductor fabricated devices with $\sim$1000's or more actuators and potentially small unit costs after initial (and possibly substantial) investment in research and development.\cite{2001SPIE.4561..147K}  The NSF-funded Center for Adaptive Optics promoted the development of MEMS technology in partnership with industry. Partially out of this effort, startup company Boston Micromachines emerged as a provider of MEMS devices for astronomy applications including NASA suborbital missions and ground-based observatories. ATI award 0906060/Baranec supported deployment of such a MEMS deformable mirror for the Robo-AO adaptive optics system on the robotic 1.5-m Palomar telescope.\cite{2012SPIE.8447E..04B,2014SPIE.9148E..12B}  Robo-AO has been used for 46 refereed publications by 32 unique first authors; results of these studies include AO imaging of nearly all of the Kepler candidate exoplanet hosts\cite{2018AJ....155..161Z} and all known stars within 25 pc observable from Mt. Palomar.\cite{BaranecPrivateCommunication}

\subsubsection{Wavefront Sensors - Keck AO}

The first NGS and LGS AO systems on an 8-10-m class telescope were deployed on the Keck~II telescope of the W.~M.~Keck Observatory.\cite{2000PASP..112..315W} Keck AO has been continually developed with substantial support from the W.~M.~Keck Foundation, NASA and LLNL; however, additional support has come from several NSF programs. Keck AO has been supported by several ATI awards, including 1611623/Wizinowich, which supported the near IR pyramid wavefront sensor.\cite{2018SPIE10703E..1ZB} 

Advantages of pyramid wavefront sensing over the more common Shack-Hartmann method include greater sensitivity within the correction band and reduced susceptibility to aliasing.   Wavefront sensing in the IR increases sky coverage because most guide stars are brighter in the IR than in visible wavebands. Pyramid wavefront sensing has been installed at LBT, Subaru, and Magellan, as well as at Keck. 

Recently the Keck pyramid wavefront sensor demonstrated its first science results.\cite{2019AJ....158..174D} The pyramid wavefront sensor is used both with NIRC2 and also with a single mode fiber injection system (funded by the Heising-Simons Foundation) that feeds the R=38,000 NIRSPEC science instrument.\cite{WizinowichPrivateCommunication} A GPU-based Real Time Controller (RTC) was developed, and this architecture is being extended to the NSF MRI-funded Keck II AO RTC and the NSF MSIP-funded Keck All-Sky Precision Adaptive Optics (KAPA) that is currently in development. 

\subsubsection{Multi-Conjugate Adaptive Optics}
MCAO uses multiple wavefront sensors and multiple deformable mirrors to achieve wide-field correction.\cite{2018ARA&A..56..277R} The region of sky over which turbulence can be considered uniform is described by the isoplanatic angle, for which a typical value may be 30$''$ at observing wavelength of 2~$\mu$m. Anisoplanatism introduces non-uniformity to the point spread function across the corrected field, limits the size of the corrected field, and effectively limits the sky coverage of AO. These restrictions motivate the use of multiple LGS and corresponding optics in GLAO, MOAO and MCAO.

Although it was first proposed in 1975\cite{1975ApJ...198..605D}, the first on-sky demonstration of MCAO was not accomplished until 2008, by the ESO supported Multi-Conjugate Adaptive Optics Demonstrator (MAD) on the VLT\cite{2008SPIE.7015E..0FM}. This was followed in 2011 by the Gemini multiconjugate adaptive optics system (GeMS) at the Gemini South telescope, which was the first system to demonstrate MCAO using multiple sodium LGS's\cite{2014MNRAS.437.2361R}. The thirty years from genesis to on-sky implementation of MCAO illustrate the technical challenges that needed to be overcome to make this idea a reality, and of the long time that it can take to develop technology for astronomy. These first generation MCAO systems will pave the way for systems that are planned for the 30-m class Extremely Large Telescope (ELT)s, including the Giant Magellan Telescope (GMT), Thirty Meter Telescope (TMT), and the European Extremely Large Telescope (ELT). 

\begin{figure}
   \begin{center}
   \begin{tabular}{c}
   \includegraphics[width=3.5in]{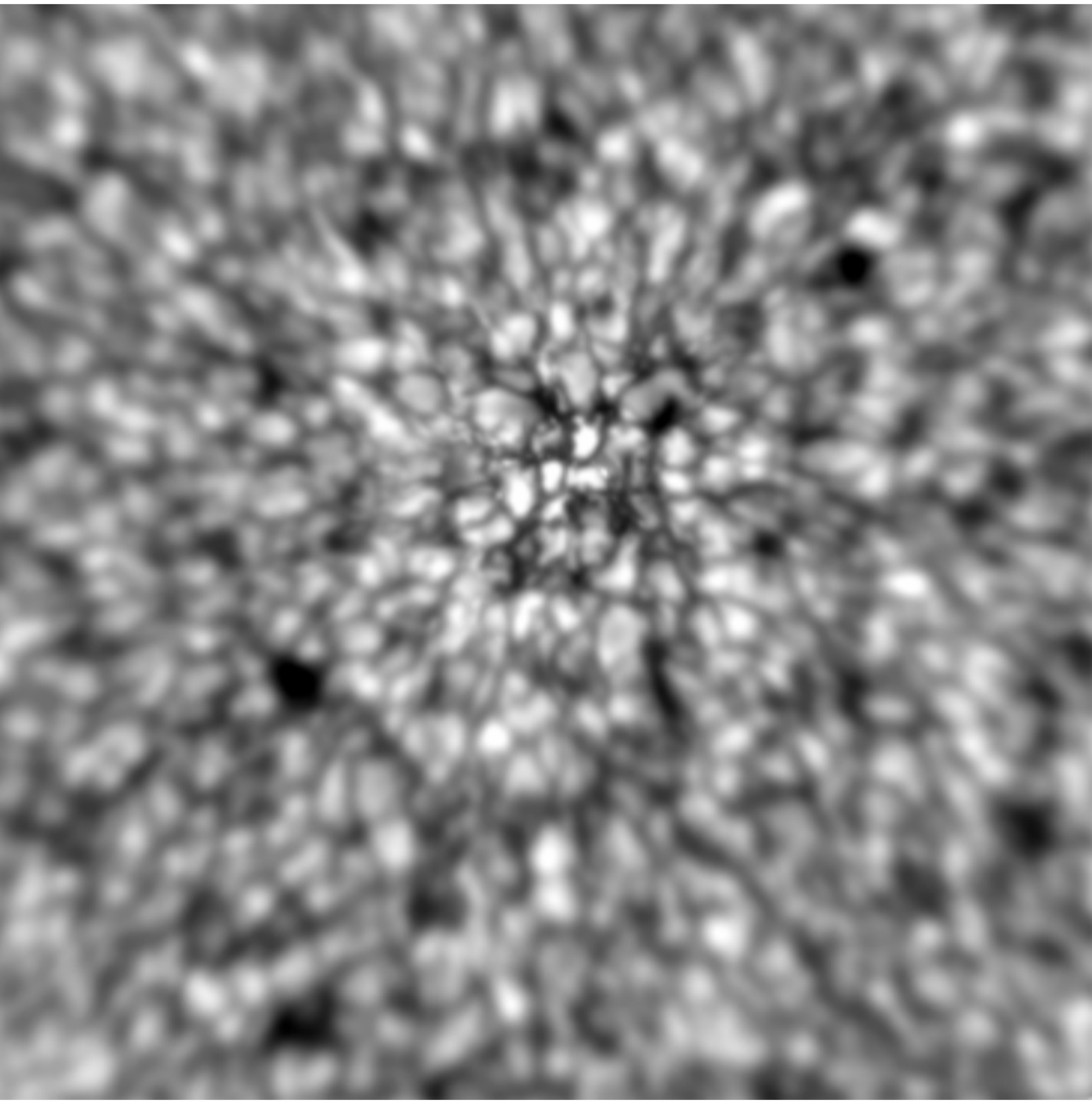}
   \includegraphics[width=3.5in]{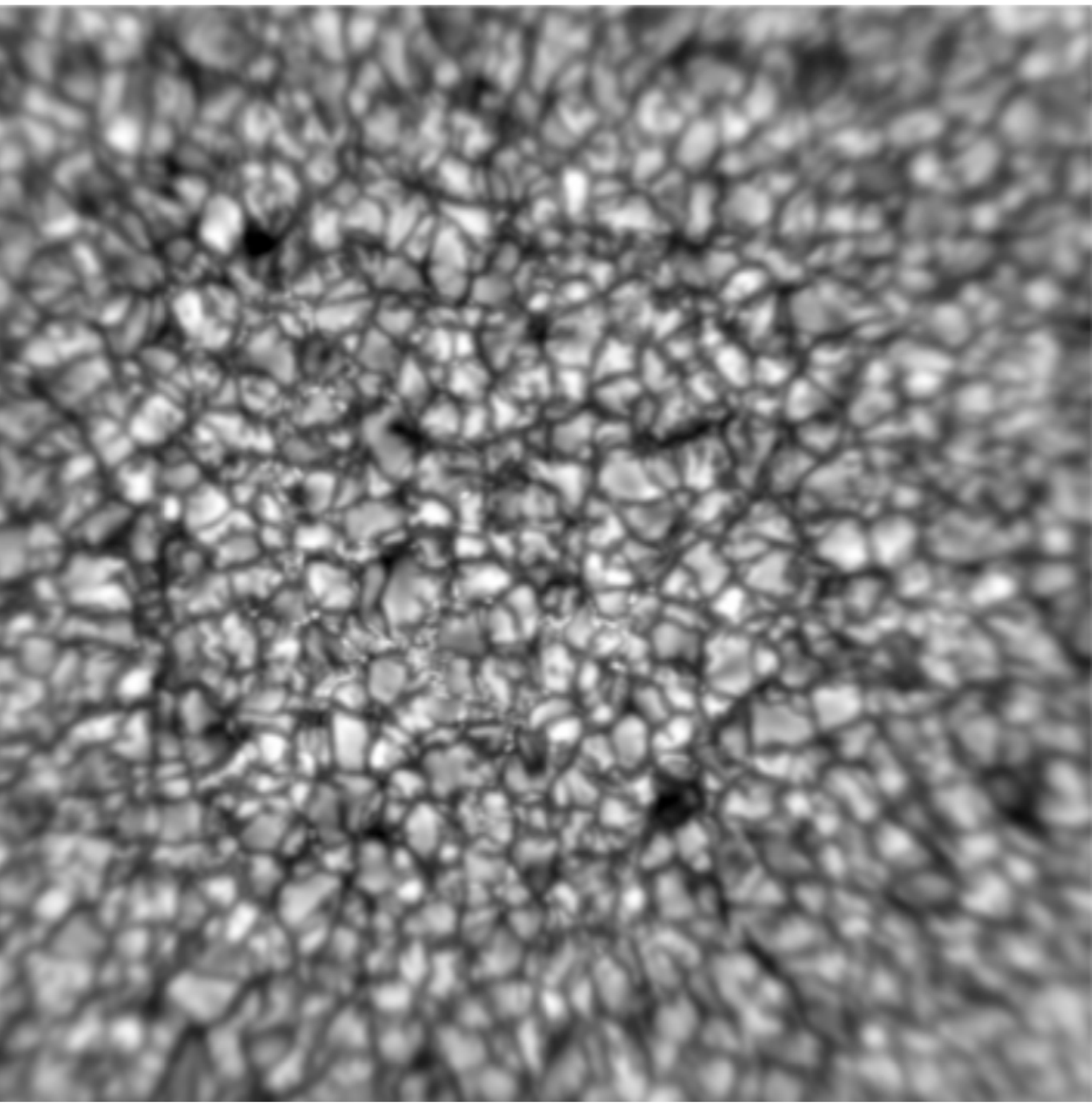}
   \end{tabular}
   \end{center}
   \caption[SolarMCAO] 
   { \label{fig:SolarMCAO} 
Images of the sun taken with Adaptive Optics (AO) at the Big Bear Solar Observatory. Observations at 0.705 $\mu$m wavelength; $50'' \times 50''$ field of view. Images taken on October 27, 2017. {\bf (left panel)} conventional AO {\bf (right panel)}.  Multi-Conjugate Adaptive Optics (MCAO). Images courtesy of Phil Goode, New Jersey Institute of Technology.}
   \end{figure}

Recently, MCAO was first demonstrated for solar imaging at the Big Bear Solar Observatory as part of an ATI supported research program (1710809/Goode).\cite{2017A&A...597L...8S}  Figure \ref{fig:SolarMCAO} is a result of this research; it illustrates the use of AO for improving solar observations, and of the wider field of correction that is enabled by MCAO. Images were made at 0.705~$\mu$m wavelength (corresponding to TiO spectral line) and have $50'' \times 50''$ field of view. Using three deformable mirrors, this MCAO system corrects more than three times the field of conventional (single conjugate) AO. This project is a pathfinder for instrumentation intended for the 4-m Daniel K. Inouye Solar Telescope. 

%

\subsubsection{Extreme Adaptive Optics}
ExAO seeks to achieve high contrast at short separations from a bright star for exoplanet searches.\cite{2018ARA&A..56..315G} Such systems must be used with bright NGS and are incompatible with LGS due to focal anisoplanatism. Science targets are completely within an isoplanatic angle, and the relevant performance metrics are contrast ratio as a function of separation between the bright star and adjacent, faint science target (\eg exoplanet). 

The first ExAO system to yield on-sky results was the ATI supported (1007046/Dekany) PALM-3000 instrument deployed at the Palomar Hale telescope.\cite{2013ApJ...776..130D} The optical design of this instrument uses two Xinetics piezoelectric deformable mirrors in a ``woofer-tweeter'' configuration to compensate for low and high spatial frequency wavefront deformations separately. PALM-3000 is designed to achieve contrast ratio of $10^{-7}$ at 1$''$ separation. The MagAO instrument on the Magellan/Clay telescope was partially supported by ATI (1206422/Close),\cite{2013ApJ...774...94C}  and its upgrade to an ExAO system is supported by an active ATI award (1506818/Males).  

Other ExAO systems include the Spectro-Polarimetric High-contrast Exoplanet Research (SPHERE) instrument for the VLT,\cite{2008SPIE.7014E..18B} the Subaru Coronagraphic Extreme AO (SCExAO) for the Subaru telescope,\cite{2009SPIE.7440E..0OM} the Gemini Planet Imager built for the Gemini Observatory.\cite{2014PNAS..11112661M} An overarching aim of ExAO efforts is the direct imaging of Earth-like exoplanets, which may require contrast ratios of $10^{-10}$ at 0.1$'''$ separation in visible wavelengths.\cite{2007Natur.446..771T} Such extraordinary precision may be beyond the reach of the current generation ExAO systems on 8-m class ground-based telescopes. However, such contrast ratios may be achievable in a space-based system or from the ground with the coming generation of 30-m class ELTs. 

It has been widely understood that adaptive optics will be essential for 30-m class telescopes to achieve their scientific potential; as such they have sometimes been referred to as {\it adaptive optics telescopes.} Due to the increasing severity of anisoplanatism at large telescope apertures, multiple laser guide star systems will be essential for many science applications. However, the most compelling, high profile science anticipated from the coming generation of ELTs will be the direct imaging of Earth-like exoplanets, for which ExAO is critical. 

In the coming decades, AO may rise to become an essential component of the next generation of large ground-based observatories. AO has grown considerably from its conception in the 1950's. Initial deployments occurred in the 1970's and 1980's. In the 1990's and 2000's AO matured into a specialized sub-discipline of astronomy. During the 2010's AO broadened and diversified to meet the needs of particular science cases. ATI support played an essential role throughout this development. 

\subsection{High Precision Radial Velocity (HPRV) Measurements of Exoplanets}

\subsubsection{HPRV spectrographs}
High precision radial velocity measurements of stars provides a key method for indirectly finding and studying exoplanets.  The wobble of a star due to its gravitational interaction with an orbiting exoplanet can be measured from the doppler shift of stellar spectral lines.\cite{2016PASP..128f6001F} Measuring these doppler shifts demands precise references for calibration. The radial velocities of exoplanet host stars may be on the order of 10's of cm s$^{-1}$, and the corresponding fractional frequency shift to be measured is of the order $\Delta f / f \sim3\times10^{-10}$ ($\Delta rv = c \Delta f / f$). 

ATI has supported such high resolution spectroscopy, particularly for exoplanet studies.  This field was pioneered from an ATI award (8919634/Marcy), which used an iodine reference cell for wavelength calibration\cite{1992PASP..104..270M}.  This method had the advantage of not requiring a special purpose spectrograph. It was adopted by others, including the HIRES instrument at the W. M. Keck Observatory\cite{1994SPIE.2198..362V}. The iodine method has a single-measurement precision floor of 1-2 m s$^{-1}$, but the superposition of a dense forest of iodine lines precludes corrections for stellar noise with new statistical approaches.\cite{FischerPrivateCommunication}

A competing approach for high precision radial velocity measurements, the cross correlation method, uses a special purpose fiber-fed spectrograph. Because of the fiber, the spectrograph can be isolated and temperature controlled, and the resulting stability has proved to be critical. Furthermore, the iodine-free spectra offer the opportunity for use of statistical methods to disentangle photospheric velocities from center of mass radial velocities. Wavelength calibration in the fiber-fed spectrographs was initially accomplished via simultaneous reference from a thorium-argon lamp. The European-funded ELODIE spectrograph\cite{1996A&AS..119..373B} at Observatoire de Haute-Provence, France used this approach to make the first detection of a Jupiter-like exoplanet around a solar-type star, 51~Pegasi\cite{1995Natur.378..355M}.  Michel Mayor and Didier Queloz at the University of Geneva shared the 2019 Nobel Prize in Physics for exoplanet research, especially as enabled by ELODIE. 

The first dedicated radial velocity spectrograph was the European-funded High Accuracy Radial Velocity Planet Searcher (HARPS) on the European Southern Observatory (ESO) 3.6-m telescope at La Silla\cite{2003Msngr.114...20M,2002Msngr.110....9P}.  HARPS achieved single measurement precision of 1 m s$^{-1}$. This instrument demonstrated the importance of vacuum-enclosed, temperature-stabilized operation, and it has been one of the leading instruments in this field since 2005.  Thanks to the improvements in the precision of wavelength calibration funded by ATI, the new NSF MRI-funded EXPRES instrument is reaching 0.20 m s$^{-1}$ single measurement precision.\cite{FischerPrivateCommunication} 

\subsubsection{Laser Frequency Combs}
In the past two decades, laser frequency combs have emerged as a critical new technology for HPRV measurements.  A frequency comb provides a spectrum of regularly spaced, narrow emission lines at known wavelengths that serve to calibrate spectroscopic line measurements to high precision. Early work on laser frequency combs used mode-locked lasers. Such lasers excite a multitude of standing wave optical modes while preserving a phase relationship between the modes (mode-locked). The resulting time domain response of the laser corresponds to coherent pulses of light; the widest bandwidth lasers may have durations of order femtoseconds; however, the frequency response is mainly of relevance for laser frequency combs. Absolute calibration of such combs was first demonstrated in 2000 using a technique called self-referencing.\cite{2000Sci...288..635J,2000PhRvL..84.5102D} In this method, the output frequency range must be broadened to roughly an optical octave (factor of two in frequency), which was first made possible by using frequency broadening photonic crystal fibers. With self-referencing, laser frequency combs could be calibrated against microwave time-frequency standards that are among the most precise known and are the basis for atomic clocks. 

Such laser frequency combs have found widespread application in optical metrology and other applications and are now commercially available. The 2005 Nobel Prize in Physics was jointly awarded to John L. Hall at JILA (a joint institute of The University of Colorado Boulder and the National Institute of Standards \& Technology, NIST) and Theodor W. H{\"a}nsch at the Max Planck Institute of Quantum Optics (MPQ) in Germany for pioneering work in this field. Subsequent research from both the JILA/NIST and MPQ groups impacted astronomy, and some research that grew out of the JILA/NIST group was supported by ATI.

One of the first applications of laser frequency combs to astronomy was accomplished by a European collaboration. An erbium-doped fiber laser frequency comb was used in conjunction with a portable atomic clock to obtain measurements of the sun with 9~m s$^{-1}$ doppler precision.\cite{Steinmetz2008} In the same year, a collaboration between the Harvard-Smithsonian Center for Astrophysics (CfA) and MIT demonstrated in the laboratory a laser frequency comb with the spectral line spacing and stability suitable for astrophysical spectrograph calibration with 1 cm s$^{-1}$ precision.\cite{2008Natur.452..610L} Subsequently, laser frequency combs were deployed at the HARPS instrument where they demonstrated substantial improvements in the repeatability and absolute calibration over the thorium-argon method.\cite{2012Natur.485..611W}

Frequency combs for IR astronomy grew out of the JILA/NIST research group in collaboration with Penn State University; this ongoing research was supported by a number of ATI awards. The IR waveband was chosen to target M dwarfs, which have properties favorable to detecting exoplanets. Use of commercial laser technology made for telecommunications applications simplified the instrument design. On-sky demonstrations using an erbium doped fiber laser were made at the Hobby Eberly Telescope.\cite{2012OExpr..20.6631Y}

Subsequently, this collaboration adopted electro-optic modulation, which is another approach to comb generation than the original femtosecond mode-locked lasers. Electro-optic frequency combs have been around since the 1960s but recently became interesting as high precision comb generators due to the advent of low-noise high frequency RF oscillators and low-driving voltage high-power handling electro-optic modulators.\cite{doi:10.1002/lpor.201300126} They are simple and robust, and permit electrical tuning of the repetition rate and central frequency. Using an electro-optic frequency comb, the NIST/Penn State collaboration recently demonstrated $\sim 1.4$~m s$^{-1}$ radial velocity precision on-sky in the IR, rivaling the precision of optical waveband HPRV measurements.\cite{2019Optic...6..233M} Recently Penn State was awarded a NASA contract to provide an HPRV system for the NEID instrument on the WIYN telescope at Kitt Peak Observatory.

The CfA group has pursued laser frequency combs in the optical wavebands, supported by several ATI awards. These awards along with private and institutional funds supported the development of a Ti:sapphire based, mode-locked laser frequency comb for the HARPS-N instrument at the TNG telescope in the Canary Islands.\cite{2014SPIE.9147E..8NL,2018JLwT...36.5309R} Recently this system was used to study radial velocity measurements of the sun due to planets in the solar system, and to better understand and characterize stellar jitter, which has emerged as a critical source of systematic error in HPRV measurements.\cite{2015ApJ...814L..21D,2019MNRAS.487.1082C} 

The current state of the art allows radial velocity measurements of astronomical objects with precision $\sim$0.8 -- 1~m~s$^{-1}$; however, the signal from Earth-like habitable zone exoplanets is expected to be closer to $\sim$10 cm s$^{-1}$\cite{2016PASP..128f6001F}.  As noted above, the required frequency precision is within the range of laser frequency combs demonstrated in the laboratory;\cite{2008Natur.452..610L} therefore there are not believed to be fundamental technical obstacles to achieving this precision on-sky. In addition, ATI supported laser frequency comb research has been devoted to improving the robustness and duty cycle of comb operation; both the NIST/Penn State and the CfA efforts have made substantial gains in robustness of laser frequency comb systems operating at a telescope. 

Technological advances continue to drive laser frequency comb development for astronomy. A third approach toward comb generation makes use of optical microresonators, which confine light to a small volume to enhance intensity and take advantage of nonlinear optical effects. A microresonator comb offers the potential to miniaturize laser frequency comb components onto a single silicon chip which could be robustly packaged for long-term, unattended operation.\cite{2011Sci...332..555K}  A recent ATI award (1908231/Vahala) pursues development of this technology. Thus in the two decades since its initial demonstration, laser frequency comb technology has advanced tremendously, and yet it is still an area of active research. The coming years may see continued maturation and dissemination of this technology for astronomy.

\subsection{Infrared Detectors}

IR detectors were primarily developed for military and commercial applications and then applied to astronomy\cite{2007ARA&A..45...43L}.  The most successful detectors to date have been based on mercury cadmium telluride (HgCdTe) semiconductor, for which development began in the 1980's.  Teledyne Imaging Sensors (formerly Rockwell Scientific Company) developed processes of HgCdTe growth by liquid phase epitaxial deposition of a Sapphire substrate - the Rockwell designation was designated as a Producible Alternative to CdZnTe Epitaxy (PACE), since molecular beam epitaxy was not available for astronomy applications. These detectors had high dark current due to the lattice mismatch between the HgCdTe and the Sapphire substrate. Rockwell was not able to measure dark current down to astronomical levels; this was done by University of Hawaii researchers to a level orders of magnitude lower than the Rockwell measurements.\cite{HallPrivateCommunication}

Independently, Raytheon Vision Systems developed processes of HgCdTe growth on silicon and CdZnTe substrates.\cite{Reddy2011} Potential advantages of the latter approach include substantially reduced cost by using established silicon processing technology.

Most of IR astronomy would not exist without NASA and NSF support.  NASA substantially funded HgCdTe detector development;  such detectors were used in the NICMOS and WFC3 instruments aboard the {\it Hubble Space Telescope}. 
Later generation HAWAII 2RG detectors are installed in the {\it James Webb Space Telescope}, ESA will launch 16 of these detectors on {\it Euclid} as well as on ground-based instruments.  There may be 30-50 HAWAII 2RG detectors deployed at ground-based observatories worldwide and the NASA WFIRST mission is base lined with 18 HAWAII 4RG-10 arrays.\cite{HallPrivateCommunication}

%
%
   \begin{figure}
   \begin{center}
   \begin{tabular}{c}
   \includegraphics[width=6.5in]{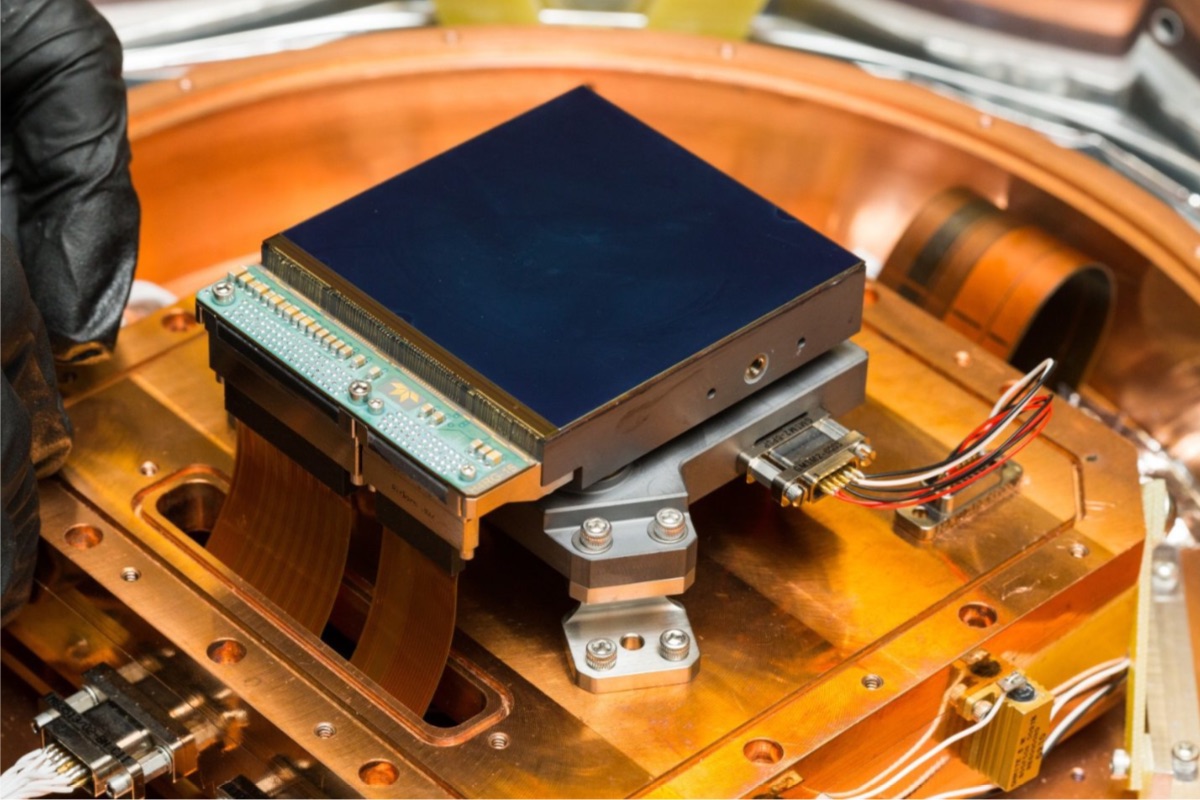}
   \end{tabular}
   \end{center}
   \caption[HAWAII4RG] 
   { \label{fig:HAWAII4RG} 
HAWAII 4RG IR detector developed with support from NASA and NSF/ATI. Figure from Hall, D.~N.~B.~\etal (2016)\cite{2016SPIE.9915E..0WH}. Reproduced with permission.}
   \end{figure} 

ATI awardees partnered with the industrial collaborators to characterize devices and tailor them to the needs of astronomy.  The largest single ATI award supported development of the HAWAII 4RG mosaic camera (0804651/Hall; \$7M, made with American Recovery \& Reinvestment Act funds) in collaboration with Teledyne, see Figure~\ref{fig:HAWAII4RG}.   Results of 0804651/Hall have been reported in conference proceedings\cite{2016SPIE.9915E..0FZ,2016SPIE.9915E..0WH} (excluded from the automated ADS search mentioned above).  Similarly, ATI awards 1207827/Figer and 1509716/Figer constituted substantial investments in the competing Raytheon technology.  At present the Raytheon detectors have difficulty reaching sufficiently low dark current\cite{Hanold2012}.  However, if further development can meet the dark current specification, Raytheon detectors could be transformative.

Traditional IR detectors are charge-integrating devices, and they are constrained by read-noise in fast readout applications.  As an alternative, Linear mode Avalanche Photo-Diode (LmAPD) arrays have been developed with support from ATI (1106391/Hall) and other funding sources in collaboration with industrial partner Leonardo (formerly Selex ES).  The Selex Advanced Photodiode HgCdTe Infrared Array (SAPHIRA) has emerged as a leading detector for near-infrared wavefront sensing in AO.\cite{Hall2016b,Atkinson2018}  On-sky tests with a near IR pyramid wavefront sensor and SAPHIRA device were recently accomplished at the Subaru SCExAO system using a camera supported by NSF\cite{2019PASP..131d4503L}, and on the Keck II telescope, as funded by a separate ATI investigation (1611623/Wizinowich).\cite{2018SPIE10703E..1ZB} Low read-noise, fast time response detectors will be crucial for pushing the limits in exoplanet studies.\cite{2015PASP..127..890J}.

\subsection{Microwave Kinetic Inductance Detectors (MKIDS)}

Another promising technology for low read-noise, fast time response detectors are Microwave Kinetic Inductance Detectors (MKIDs)\cite{2003Natur.425..817D}.  Unlike CCDs or photodiodes, these detectors rely upon thin superconducting films that are arranged in a microwave resonant circuit.  Incident photons break Cooper pairs and generate quasiparticles which change the complex impedance and hence the resonance of the circuit.  High quality factor resonant circuits (operating at cryogenic temperatures) can be sensitive to individual photons.  Individual detector elements can be made with unique microwave resonance frequencies and multiplexed into large arrays with simple readouts.  This technology has broad application in addition to optical-IR astronomical detectors\cite{2012OExpr..20.1503M}.  

%
   \begin{figure}
   \begin{center}
   \begin{tabular}{c}
   \includegraphics[width=6.5in]{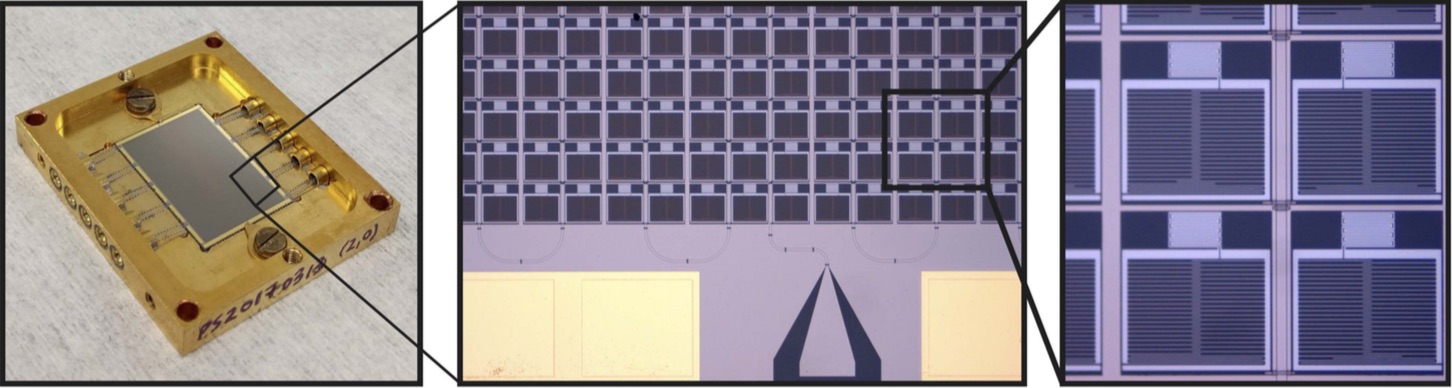}
   \end{tabular}
   \end{center}
   \caption[MKID] 
   { \label{fig:MKID} 
Microwave Kinetic Inductance Detector (MKID) from the DARKNESS instrument supported by ATI. {\bf (left panel)} MKID device mounted in package {\bf (middle panel)}.  Detail image showing several rows of pixels, transmission line and bondpad for one feedline. {\bf(right panel)} Detail of several MKID pixels; densely meandered patches at the top of each pixel are the photosensitive inductors, and the large sparse sections are the interdigitated capacitors used to tune each MKID pixel to a unique resonant frequency. Figure from Meeker \etal (2018)\cite{2018PASP..130f5001M}. \copyright ~The Astronomical Society of the Pacific. Reproduced by permission of IOP Publishing. All rights reserved.}
   \end{figure} 

The first application of an MKID detector to ground-based, optical-IR astronomy was the Array Camera for Optical to Near-IR Spectrophotometry (ARCONS)\cite{2013PASP..125.1348M,2015ApJS..219...14V}.    Its successor, the DARK-speckle Near-infrared Energy-resolving Superconducting Spectrophotometer (DARKNESS) was supported by an ATI award (1308556/Mazin), see Figure~\ref{fig:MKID}; it was recently installed behind the PALM-3000 adaptive optics system and the Stellar Double Coronagraph at Palomar Observatory\cite{2018PASP..130f5001M}.  In subsequent research from this same research group that was supported by NASA, a 20 kilo-pixel MKID array was fabricated for the MKID Exoplanet Camera (MEC) intended for use at the Subaru telescope.\cite{2017OExpr..2525894S} The sensitivity and fast time response of MKIDs makes them amenable to ExAO and speckle imaging/nulling applications, which are the subjects of active ATI supported investigations.

\subsection{Multi-Object Spectroscopy}

Multi-object spectroscopy has been developed through ATI awards.  The FLoridA Mulit-object Infrared Grism Observational Spectrometer (FLAMINGOS) system was the first cryogenic, multi-object spectrograph, and it was developed through an ATI award (9731180/Elston).  This instrument accomplished the technical challenges of masking the focal plane to allow selecting specific objects for study in a manner that is reconfigurable, dispersing the light and finally detecting it, all while operating at cryogenic temperature at the telescope\cite{2003SPIE.4841.1611E}.  FLAMINGOS cryostats were a significant innovation, as well as adaptation of the optical imaging spectrograph approach to cryogenic infrared optical materials.  Selectable slit masks for multi-object target selection at the focal plane were laser-machined and placed in a rapidly thermal-cycled fore-dewar while the rest of the instrument could be kept cold for long durations.  More recent designs use dynamically reconfigurable slit units.  

Although the particular approach of a selectable machined mask used in FLAMINGOS was not widely adopted, this instrument served as a proof of concept.  It demonstrated that such a complex system could be made to work at a telescope and it undoubtedly inspired other efforts.  Although this award was acknowledged in the literature only five times, it nevertheless had a profound effect.  Subsequent multi-object spectrographs include the Multi-Object Infrared Camera and Spectrograph (MOIRCS) for the Subaru telescope\cite{2008PASJ...60.1347S},  FLAMINGOS-2 on Gemini\cite{2012SPIE.8446E..0IE}, EMIR on Gran Telescopio Canarias\cite{2016SPIE.9908E..1JG}, Multi-Object Spectrometer For Infra-Red Exploration (MOSFIRE) on Keck\cite{2012SPIE.8446E..0JM}, MMT and Magellan Infrared Spectrograph (MMIRS), which was based explicitly on the FLAMINGOS design\cite{2012PASP..124.1318M}, the LBT NIR-Spectroscopic Utility with Camera and Integral-Field Unit for Extragalactic Research (LUCIFER)  on the Large Binocular Telescope\cite{2010SPIE.7735E..1LA}. Also notable is the K-band Multi-Object Spectrograph (KMOS) on the Very Large Telescope, which is really a multi-IFU spectrograph that does not have slit masks\cite{2006SPIE.6269E..1CS,2013Msngr.151...21S}.  

\subsection{Optical Interferometry}

Interferometry coherently combines light from multiple apertures or telescopes to synthesize data for a single object. Optical interferometry provides exquisite resolution that is unobtainable with filled aperture telescopes; modern optical interferometers achieve limiting resolutions in the sub-milliarcsecond regime. 
Such instruments have been used to measure the diameters, distances and/or other properties of solar system objects, individual stars, and recently even active galactic nuclei. 

Optical interferometry was first demonstrated in the 1920's as a method to determine stellar properties\cite{1920ApJ....51..263A,1921ApJ....53..249M}. Interferometry was further developed for astronomy in the late 1970's and 1980's, which led to instruments such as the Sydney University Stellar Interferometer (SUSI)\cite{1994SPIE.2200..231D} and the Mark III interferometer\cite{1992ApJ...384..624P,1993ApJ...413L.129P,1995AJ....110..376H,1987AJ.....93.1280S}. The current generation of interferometers includes the Navy Precision Optical Interferometer (NPOI)\cite{1998ApJ...496..550A}, the Palomar Testbed Interferometer (PTI; decommisioned)\cite{1999ApJ...510..505C}, the Center for High Angular Resolution Astronomy array (CHARA)\cite{2005ApJ...628..453T}, the (decommisioned) Keck Interferometer\cite{1999ASPC..194..256B}, Magdalena Ridge Optical Interferometer (MROI; not yet operational)\cite{2018SPIE10701E..06C} and especially the Very Large Telescope Interferometer, VLTI, experiment GRAVITY \cite{2008SPIE.7013E..2AE,2017A&A...602A..94G}. GRAVITY is supported by the European Southern Observatory whereas the major US based initiatives include NPOI, MROI and CHARA. 

CHARA has been supported through six NSF/ATI grants beginning in 1992 and a current MSIP award, as well as through substantial private foundation support. ATI has been ``essential in many ways" for CHARA.\cite{tenBrummelaarPrivateCommunication} These grants have supported the use of adaptive optics in CHARA, which improves the sensitivity of interferometric images (as opposed to improving the resolution, which is the usual provenance of adaptive optics). They have also supported development of the Michigan Young Star Imager at CHARA (MYSTIC)\cite{2018SPIE10701E..22M}, which is a K-band cryogenic 6-beam combiner to image disks around young stars in order to study planet formation. An MSIP award supports public access to CHARA in the amount of 50-75 nights per year; observing time is administered through the National Optical Astronomy Observatories.

In addition to measuring diameters of bright stars, interferometry has determined orbits of spectroscopic binaries with high precision \cite{1998AJ....116.2536H,2001AJ....121.1623H,2002AJ....124.1716T,2010AJ....140.1623M,2010AJ....139.2308F}, studied the mass-radius relation of dwarf stars\cite{2006ApJ...644..475B}  and studied the effects of rotation on stellar surfaces\cite{2005ApJ...628..439M,2006ApJ...637..494V,2012ApJ...761L...3M}. Future science applications include the study of AGN cores, young stellar objects and exoplanet host stars. 

In the 1970's interferometry was extended into the mid-IR waveband. This advance was enabled by the availability of IR detectors with fast time response (less than one nanosecond) for use as mixing elements in heterodyne detectors\cite{1973NPhS..241...32G}.  A two-element interferometer was deployed at Kitt Peak\cite{1974PhRvL..33.1617J,1988ESOC...29..867D} which studied planet forming disks. Another instrument at Mount Wilson Observatory was the product of ATI support (9016474/Townes and 9119317/Townes); this Infrared Spatial Interferometer (ISI) provided the first detailed, resolved observations of dust shells around late type stars\cite{1994AJ....107.1469D}.

Early implementations used amplitude interferometry, however in the 1950's, Hanbury Brown and Twiss exploited correlations in intensity (``intensity interferometry") to interferometrically measure diameters of bright stars\cite{1956Natur.177...27B,1957RSPSA.242..300B}.  Meanwhile advances in fast electronics and photon detectors have enabled revisiting earlier intensity interferometry techniques \cite{2010SPIE.7734E..1DL,2012MNRAS.424.1006N}. The European Cerenkov Telescope Array (CTA)\cite{2016SPIE.9907E..0MD} and the US-led ATI funded effort uses the VERITAS array (1806262/Kieda)\cite{2019ICRC...36..714K}.

\subsection{Radio Wave Digital Signal Processing}

Ever since the discovery of celestial radio emission by Karl Jansky in the 1930's at Bell Laboratories\cite{1933PA.....41..548J}, astronomers have benefited from technology to detect and process radio waves. Radio astronomy entered the digital era in the 1960's with the advent of the first astrophysical digital radio spectrum analysis.\cite{1963Natur.200..829W} Both single dish radio antennas and antenna arrays have benefitted from tremendous progress in signal processing. There are diverse approaches to processing of contemporary radio astronomical data. Some typical functions that are important for arrays may include: (1) digitizing the analog sky signal (2) channelizing this digitized signal into discrete frequency bins and (3) combining signals from multiple antennas. This last function may consist of weighted addition of antenna signals, known as beamforming, or as correlation (\ie multiplication) of antenna signals. A common design pattern consists of dividing the ``front end'' digitization and channelization functions from the ``back end'' higher level tasks that may be performed on a different compute architecture.\cite{2016JAI.....502002P} Front end systems may use Field Programmable Gate Arrays (FPGAs), which are reconfigurable integrated circuits. Back end systems may employ Graphical Processing Units (GPUs), which are massively parallel compute engines. In general, the specialized computational methods of radio astronomy, such as those outlined above, add a considerable burden of complexity and a barrier to learning for the students and early career researchers who may ultimately build these systems.

The Collaboration for Astronomy Signal Processing and Electronics Research (CASPER) has provided open source technology and a supportive community for making this technology available for more than a decade. CASPER hardware and software powers more than 45 radio astronomy instruments worldwide.\cite{2016JAI.....541001H}  CASPER has been supported by seven ATI grants from 2002-2019 that have included awards for highly regarded conferences / workshop tutorials in FPGA and GPU programming. CASPER hardware included the once popular Reconfigurable Open Architecture Computing Hardware (ROACH) series of standalone FPGA processing boards. CASPER powered spectrometers include the Allen Telescope Array Fly's Eye radio transient search,\cite{2012ApJ...744..109S} the Green Bank Ultimate Pulsar Processing Instrument (GUPPI)\cite{2008SPIE.7019E..1DD} and the Versatile GBT Astronomical Spectrometer (VEGAS)\cite{2014PASA...31...48C}, which is the workhorse back end system for the GBT, and many others. CASPER based correlators and beamformers have been used for the Precision Array for Probing the Epoch of Reionization (PAPER)\cite{2010AJ....139.1468P}, Hydrogen Epoch of Reionization Array (HERA)\cite{2016icea.confE...2D} and Large Aperture Experiment to Detect the Dark Ages (LEDA)\cite{2015JAI.....450003K}, and the Focal L-band Array for the GBT (FLAG)\cite{2018AJ....155..202R} discussed below, as well as the Smithsonian Astrophysical Observatory's Submillimeter Array (SMA) Wideband Astronomical ROACH2 Machine (SWARM)\cite{2016JAI.....541006P}, the Breakthrough Listen\cite{2018PASP..130d4502M} and other projects to search for extratrerrestrial life. CASPER hardware was used for the Very Long Baseline Array (VLBA)\cite{2010ivs..conf..396N} and it was subsequently used for the millimeter wavelength Very Long Baseline Interferometry (VLBI) as part of the Event Horizon Telescope.\cite{2015PASP..127.1226V} As of 2017, this hardware enabled VLBI data recording at 64 gigabits per second, which is nearly five orders of magnitude improvement over the original VLBI recording rates circa 1969. One trend that CASPER has encouraged in recent years is the routing of data via Ethernet switches, a so-called ``packetized" approach.

CASPER hardware has also been applied outside of traditional radio astronomy, notably to MKID readouts, which require similar processing functions. Examples include the ATI supported MUltiwavelength Sub/millimeter Inductance Camera (MUSIC)\cite{2012SPIE.8452E..05G} that was deployed at the Caltech Submillimeter Observatory,  a millimeter wavelength system developed at Columbia University\cite{2014RScI...85l3117M} and the optical instruments DARKNESS and MEC\cite{2018PASP..130f5001M}. A Transition Edge Sensor, discussed below, that uses CASPER hardware is the Muliplexed SQUID TES Array at Ninety Gigahertz (MUSTANG-2)\cite{2016JLTP..184..460S}, which is deployed at the GBT. Hickish \etal includes a census of CASPER deployments for radio astronomy as of 2016.\cite{2016JAI.....541001H} 

In addition to CASPER, several other general purpose digital signal processing platforms have been developed for radio astronomy. The Uniboard\cite{2010evn..confE..98S} FPGA platform was developed under a Joint Research Activity in the RadioNet FP7 European programme; the radionet consortium coordinates the European radio astronomy community.\cite{RadioNet} Several generations of digital signal processing platforms, named Redback, have been developed independently for the Australian Square Kilometer Array Pathfinder (ASKAP) array, and these platforms were partly inspired by CASPER.\cite{inproceedings}

\subsection{Radio Wave Phased Array Feeds}

The extreme faintness of celestial sources motivates telescopes with large collecting areas, which necessarily have correspondingly small fields of view. Therefore array feeds consisting of multiple antennas located at the telescope focal plane have been used to increase field of view compared to a single feed antenna. Feeds consisting of dense arrays of electrically small antennas, called Phased Array Feeds (PAFs) offer the potential to sample a larger area of the focal plane than possible with a single-pixel horn feed, and thereby increase mapping speed for surveys. Multiple beams are formed by combining signals sampled by the array elements with complex weights. PAFs were initially used for radar and satellite transmission; however in the late 1990's and early 2000's, they began to be applied to astronomy.\cite{2000SPIE.4015..308F}. 

\begin{figure}
   \begin{center}
   \begin{tabular}{c}
   \includegraphics[height=5cm]{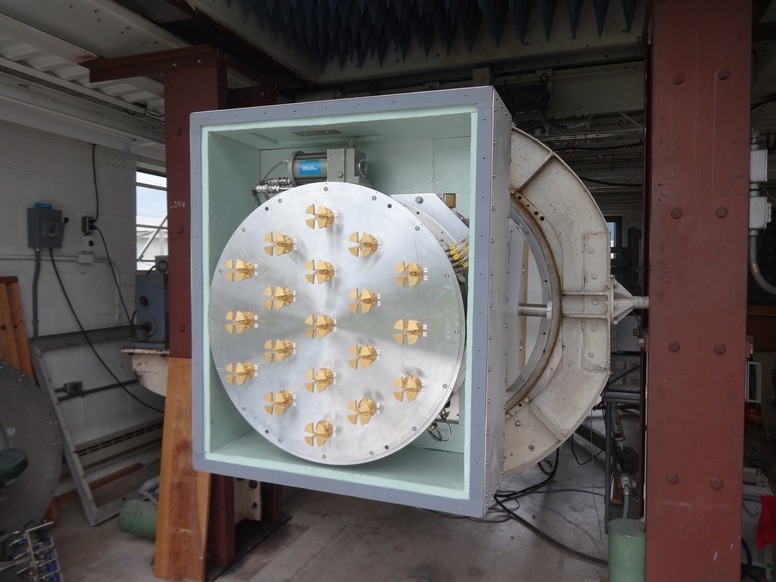}
   \includegraphics[height=5cm]{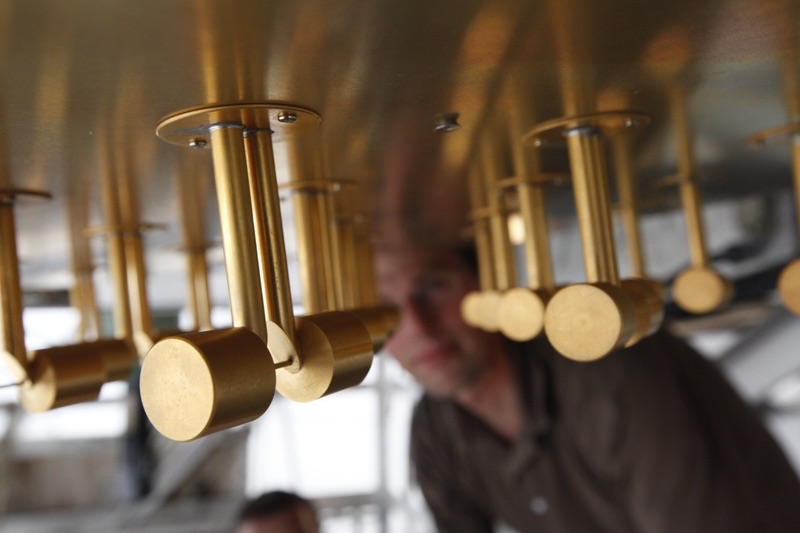}
   \end{tabular}
   \end{center}
   \caption[PAFs] 
   { \label{fig:PhasedArrayFeeds} 
Phased Array Feeds supported by ATI {\bf (left panel)} The FLAG instrument deployed at Green Bank Observatory. Image courtesy of Brian Jeffs, Brigham Young University {\bf (right panel)}.  Engineering prototype PAF at Arecibo Observatory. Image courtesy of Mark Philbrick, Brigham Young University.}
   \end{figure}

Early work was done by the National Radio Astronomy Observatory (NRAO) on the 43-m Green Bank Telescope (GBT)\cite{2000SPIE.4015..308F,2000ASPC..217...11F}, with continuing collaboration with Brigham Young University (BYU). In particular, ATI award 9987339/Jeffs supported real-time DSP-based radio frequency interference mitigation strategies that used multiple antennas for tracking and nulling sources of interference.\cite{2005ITSP...53..439J} This research grew to encompass PAFs, and it was supported by three more ATI awards as well as ten awards from other NSF programs in the 2000's and 2010's to support PAF development. These awards included support for the first millimeter-wave PAF, PHAMASS, in collaboration with University of Massachusetts, Amherst.\cite{2015isap.confE...3E} ATI supported a collaboration between BYU and West Virginia University to develop the Focal L-Band Array for the Green Bank Telescope (FLAG) instrument, see Figure~\ref{fig:PhasedArrayFeeds}, which was recently commissioned\cite{2019MNRAS.489.1709R,2019ITAP...67.3011R}, and ATI and MSIP awards are supporting BYU and Cornell in development of the Advanced L-band cryogenic Phased Array Camera for Arecibo (ALPACA). This instrument is expected to deliver the largest PAF to date for a US telescope, and the only fully cryogenically cooled (LNAs and antennas) L-band PAF in the world.\cite{2015ITAP...63.2471C,2016SPIE.9908E..5FC}
%
From the beginning, PAFs for astronomy were pursued independently by several research groups, including the National Research Council of Canada\cite{7852186}, and especially  Commonwealth Scientific and Industrial Research Organization (CSIRO) in Australia for the Parkes telescope\cite{2014PASA...31...41H} and ASTRON in the Netherlands\cite{2000SPIE.4015..420V}. Research at ASTRON included technology demonstrators that paved the way for designs of the Square Kilometer Array (SKA) and separately for LOFAR\cite{2000SPIE.4015..420V,2009IEEEP..97.1531V}. The Australian group fielded a PAF on the Australia Square Kilometer Array Precursor (ASKAP) telescope.\cite{2009IEEEP..97.1507D,2010ITAP...58.1922H,2010past.conf..648C}

Science results have begun to emerge from PAF technology development. The APERture Tile In Focus (APERTIF) instrument installed at the Westerbork Synthesis Radio Telescope (WSRT) in the Netherlands presented early results of on-sky testing\cite{2010iska.meetE..43O}. Since the discovery of Fast Radio Bursts (FRBs)\cite{2007Sci...318..777L}, the study of radio transients has taken on particular urgency. Surveys for transients and pulsars have been accomplished using PAF installed at ASKAP\cite{2016MNRAS.456.3948H,2018MNRAS.478.1784B,2019MNRAS.486..166Q} and dedicated monitoring of the Vela pulsar with the Parkes 12-m antenna and Mark I PAF \cite{2017PASA...34...27S,2017PASA...34...26D}. Notably the PAF survey with ASKAP accomplished in days a survey of comparable area (30 sq. deg) as was done in weeks-months with the Karl Jansky VLA \cite{2016ApJ...818..105M}, albeit to somewhat shallower depth.\cite{2018MNRAS.478.1784B} A pilot survey of 10 sq deg with the FLAG instrument on GBT demonstrated instrument performance on-sky although it did not report FRB or pulsar detections\cite{2019MNRAS.489.1709R}. The ALPACA PAF, for which ATI supported a concept study \cite{2016SPIE.9908E..5FC} is expected to be the most sensitive PAF to date and will discover pulsars up to an order of magnitude faster than others.\cite{2019MNRAS.489.1709R} These early results, coming nearly two decades after the initial papers on PAFs for astronomy, demonstrate the long time that it can take for technology to mature. 

\subsection{Cosmology and FIR / Microwave Detectors}

The study of the origin and evolution of the universe as a whole became the precision science of cosmology due to observations of the cosmic microwave background (CMB). This background radiation was famously discovered in 1965 at Bell Laboratories.\cite{1965ApJ...142..419P} The original discovery, as well as the later discovery, by the COBE satellite launched by NASA in 1989, of the blackbody spectrum and anisotropy of this radiation\cite{1990ApJ...354L..37M,1994ApJ...420..439M,1992ApJ...396L...1S} were awarded separate Nobel prizes. These discoveries provided crucial evidence for the Big Bang and the origins of cosmic structure. 

Subsequent ground-based and balloon-based measurements revealed acoustic peaks in the angular power spectrum of the CMB, which provided evidence of flat spacetime and supported claims of accelerated expansion as revealed by observations of Type Ia supernovae. Polarization in the CMB was  discovered\cite{2002Natur.420..772K,2004Sci...306..836R,2005ApJ...619L.127B}, and the cosmic microwave background remains an extremely active field. Community-wide plans for ambitious  ``CMB S4'' experiment to map polarization -- from gravitational lensing and potentially a primordial background of inflationary gravitational waves -- are underway\cite{2016arXiv161002743A}. 

Motivated in large part by CMB science, detectors for the FIR-microwave regime, herein defined as radiation from microns to millimeters in wavelength, or equivalently from gigahertz through terahertz frequencies, have been developed. They included coherent heterodyne receivers adapted from radio astronomy. They also included incoherent detectors, mainly bolometers that were developed separately\cite{2007ARA&A..45...43L}. 

\subsubsection{Coherent (Heterodyne) Detectors}

Key early progress on the CMB,  including its original discovery, as well as the discovery of fluctuations by COBE and the discovery of CMB polarization by ground-based observations were done with coherent detectors. Such detectors amplify the detected signal and change its frequency while preserving its phase.\cite{2004tra..book.....R}  The large amplification ($\sim 100$ dB) needed to detect faint astronomical signals makes amplifiers highly susceptible to feedback oscillation; therefore, to achieve stability, the amplified signal is mixed with a local oscillator to change its output frequency. This system forms a heterodyne receiver. Front-end amplifiers for such systems have included High Electron Mobility Transistors (HEMTs), Superconductor-Insulator-Superconductor (SIS) mixers and Hot Electron Bolometers (HEBs); but maser amplifiers have the lowest noise.

ATI award 8024119/Readhead supported development and use of a ruby maser for the 40-m telescope at Owens Valley Radio Observatory (OVRO). It was used to place early limits on CMB anisotropy.\cite{1989ApJ...346..566R} These limits were the most stringent at the time, and the results, which were published in 1989, required both both dark matter and dark energy.\cite{ReadheadPrivateCommunication} They presaged  evidence for the latter from observations of Type 1a supernovae by a decade.

\begin{figure}
   \begin{center}
   \begin{tabular}{c}
   \includegraphics[width=5in]{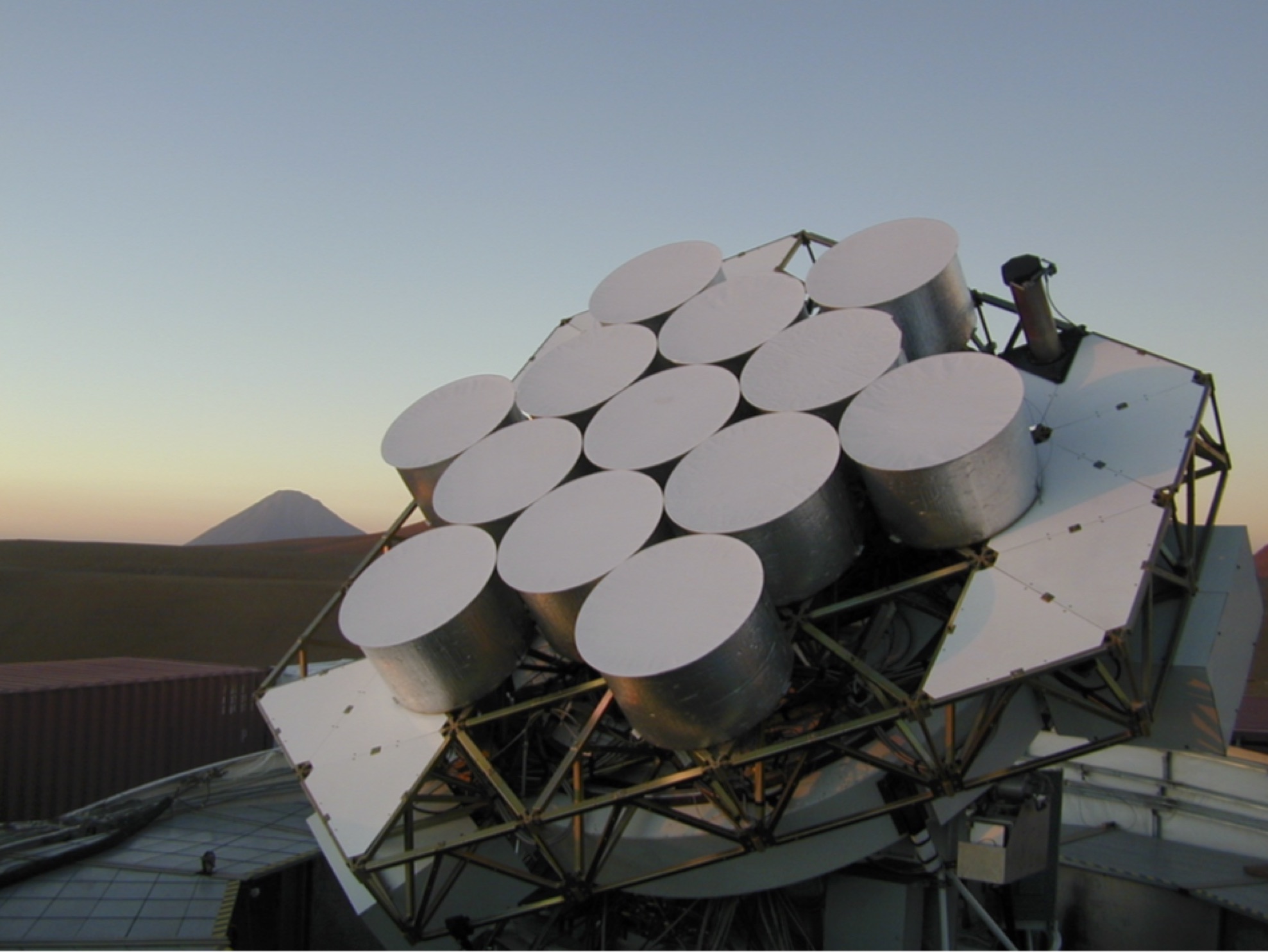}
   \end{tabular}
   \end{center}
   \caption[CBI] 
   { \label{fig:CBI} 
The Cosmic Background Imager (CBI) which measured CMB polarization and was supported by several ATI awards. Image courtesy of Anthony~C.~S.~Readhead.}
   \end{figure}

The Degree Angular Scale Interferometer (DASI) instrument, which was supported by the Center for Astrophysics Research in Antarctica (CARA, an NSF Science and Technology Center), the NSF Office of Polar Programs, and partially by ATI, measured the angular power spectrum of the CMB; these results were the most widely cited in the history of the ATI program.\cite{2002ApJ...568...38H} DASI also accomplished the first detection of CMB polarization.\cite{2002Natur.420..772K}   As mentioned above, the DASI electronics and correlator were partially supported by ATI, which principally supported a sister instrument, the Cosmic Background Imager (CBI), see Figure~\ref{fig:CBI}, through several awards between 1994 and 2002. CBI made the second measurement of CMB polarization in 2004.\cite{2004Sci...306..836R} 

%
%

ATI also supported millimeter wavelength interferometry through awards to the University of Chicago, beginning with an NSF Young Investigator award to John Carlstrom in 1992. Later ATI awards 0096913/Carlstrom and 0604982/Carlstrom supported an array of six 3.5-m telescopes and a wideband digital correlator for millimeter wavelength interferometry that were deployed at OVRO as the Sunyaev-Zel'dovich Array (SZA).  These telescopes were used to make measurements of the Sunyaev-Zel'dovich (SZ) effect on galaxy clusters for cosmology and astrophysics. 

The SZ effect is an absorption of the CMB caused by inverse Compton scattering of CMB photons by hot electrons, principally in the vicinities of clusters of galaxies.\cite{2002ARA&A..40..643C} It is important for cosmology. Measurements of this effect can be combined with other diagnostics to measure the cosmological distance scale, and the number density of SZ-detected galaxy clusters can be used to probe dark energy. The first unambiguous detection of the SZ effect was accomplished with the single dish 40-m telescope at OVRO;\cite{1984Natur.309...34B} this effort was made possible by the aforementioned ruby maser, and it was supported by an additional NSF award (8210259/Cohen). However, interferometer arrays were preferred over single dish telescopes due to their stability and spatial filtering capability. 

The SZA continued operation until 2007. By combining its microwave measurements with X-ray measurements of hot cluster gas, independent distance estimates to galaxy clusters were obtained, which provided an independent measurement of the Hubble constant, $H_0$.\cite{2006ApJ...647...25B} In 2008, these telescopes were combined with OVRO telescopes as well as telescopes from the Berkeley-Illinois-Maryland Association (BIMA) and re-deployed to the Inyo Mountains in California near OVRO to complete the Combined Array for Radio Millimeter Astronomy (CARMA). CARMA was supported by NSF through cooperative agreements until 2015. It was an important precursor to the Atacama Large Millimeter-submillimeter Array (ALMA), and it provided essential training for instrumentalists. 

ATI award 0905855/Church developed a prototype scalable 4-pixel heterodyne array based on MMIC amplifiers operating at 70-116 GHz. In a subsequent ATI award, 1207825/Church, this technology was used to build the 16-pixel Argus spectroscopic array, currently deployed on the Green Bank Telescope.

\subsubsection{Semiconductor Bolometers}

Bolometers operate on the principle that incident radiation is absorbed and thermalized within the detector. The resulting minuscule temperature change is converted by a thermistor, amplified and sensed to generate a signal\cite{Richards:2005jc,Rieke:2007fn}. Bolometers are operated at very low temperature to reduce thermal noise and improve response time. Silicon nitride has been used to thermally isolate and mechanically suspend an absorbing gold coating and a neutron transmutation doped Germanium semiconductor used as the thermistor. More recently, bolometers using a Transition Edge Sensor (TES) as a sensitive thermometer have been coupled with superconducting readout circuits to produce multiplexed arrays; they are discussed separately below.

ATI substantially impacted bolometers for millimeter-wave astronomy. Notably 9503276/Lange supported the Sunyaev-Zeldovich Infrared Experiment (SuZIE)\cite{1997ApJ...479...17H}. An innovation of this experiment was the silicon nitride (`spider web') bolometer. In the spider web geometry, the absorber is patterned into a mesh structure with the thermistor at its center. This geometry lowers the heat capacity compared to a monolithic detector. As a result, the detector has a larger throughput without loss of sensitivity, smaller cross section to contaminating cosmic rays and reduced sensitivity to microphonics\cite{1995SSRv...74..229B}. This architecture was later adopted in the {\it Herschel Space Observatory} Spectral and Photometric Imaging Receiver (SPIRE)\cite{1998SPIE.3357..404G,1998SPIE.3357..297B}, the Balloon Observations Of Millimetric Extragalactic Radiation and Geomagnetics (BOOMERANG)\cite{2000Natur.404..955D}, the Arc-Minute Bolometer Array Receiver (ACBAR)\cite{2003ApJS..149..265R}, the {\it Planck} satellite High Frequency Instrument (HFI)\cite{Lamarre:2010ck} and others. 

The single-element design then expanded to arrays in the successor to SuZIE, the Bolocam instrument \cite{1998SPIE.3357..326G,2004SPIE.5498...78H} (partially supported by NSF award 0098737/Lange). Bolocam deployed a large-scale single wafer bolometer and horn arrays, cold readout electronics and a software pipeline. Such techniques were implemented in other programs. Bolocam helped pave the way for other arrays with feedhorn-coupled bolometers including the Atacama Cosmology Telescope (ACT)\cite{2010ApJS..191..423H}, ACBAR, the Atacama Pathfinder Experiment (APEX)\cite{2006A&A...454L..13G}, the Atacama Submillimeter Telescope Experiment (ASTE)\cite{2004SPIE.5489..763E}, Herschel SPIRE and generations of instruments for the South Pole Telescope (SPT)\cite{2004SPIE.5498...11R,2011PASP..123..568C}.

ATI supported research also impacted polarization sensitive bolometers for studying the CMB. 0096778/Church supported development of the QUEST and DASI instrument (QUaD)\cite{2003SPIE.4855..227J}. QUaD pioneered the use of bolometers to detect and study polarization in the CMB\cite{2008ApJ...674...22A,2009ApJ...692.1247P,2009ApJ...692.1221H}. Prior to QUaD, the first CMB polarization measurements were done with coherent receivers, which were preferred due to their higher intrinsic polarization sensitivity. However, the QUaD bolometer system accomplished comparable polarization sensitivity to coherent receivers while also retaining intrinsic advantages of bolometers, such as far higher instantaneous sensitivity over wider bandwidths, better stability and scalability to higher frequencies\cite{2003SPIE.4855..227J}. Subsequent orbital, sub-orbital and ground-based bolometric CMB polarization experiments including the {\it Planck} satellite HFI\cite{Lamarre:2010ck}, BOOMERANG\cite{2000Natur.404..955D} and BICEP  based their receiver designs on those developed for QUaD. Numerous other groups have used or plan to use bolometers at ground-based sites to explore CMB polarization including ACTPol\cite{2010SPIE.7741E..1SN}, Advanced ACTPol\cite{2016JLTP..184...66L}, BICEP2-3\cite{2016JLTP..184..765W} and BICEP Array, CLASS\cite{2014SPIE.9153E..1IE}, Keck Array\cite{2012JLTP..167..827S}, POLARBEAR\cite{2014JLTP..176..726B}, POLARBEAR-2\cite{2014JLTP..176..719S}, Simons Observatory\cite{2019JCAP...02..056A}, SPT-Pol\cite{2012SPIE.8452E..1EA} and SPT-3G\cite{2014SPIE.9153E..1PB}. Thus polarization sensitive bolometers opened a rich and exciting area of CMB research.

\subsubsection{Transition Edge Sensors}

Detectors for studying the CMB have advanced considerably, in particular due to development of the superconducting Transition Edge Sensor (TES). A TES consists of a superconducting film operated in a narrow temperature range over which the material transitions between a normal and superconductive state.\cite{2018PhT....71h..28M} The principle of detection relies upon energy deposition in the film causing a change in electrical resistance. TES detectors can measure power (bolometers) or pulses of energy (calorimeters). Initially demonstrated in 1941\cite{1941PhRv...59.1045A}, their development advanced when they were coupled with Superconducting Quantum Interference Device (SQUID) current amplifiers, which were easily impedance matched to the low resistance of TES detectors \cite{2005cpd..book...63I}. Further technical advances included voltage biased operation\cite{1996ApPhL..69.1801L} and multiplexing the SQUID readout \cite{2006NIMPA.559..786L}. Substantial technical development occurred between 1995-2005, much of which was supported by NASA\cite{2008tdcp.workE...1B}. However, ATI award 9731200/Richards, which supported TES development at UC Berkeley, resulted in the first demonstration of a TES bolometer made entirely from photolithography\cite{1999ApPhL..74..868G,2000ApPhL..77.4040G} as well as SQUID frequency multiplexing\cite{2001ApPhL..78..371Y}. Full lithographic fabrication was a crucial step enabling the development of large arrays. Photolithography fabricated TES bolometer arrays have been fielded in numerous experiments to date; they are also being developed for near-term experiments and may be important for CMB-S4 \cite{2018JLTP..193..633H}.

\subsection{Cosmic Dawn, the Epoch of Reionization and Low Frequency Radio Experiments}

Understanding the origins of the first stars and galaxies is a great challenge of astrophysics. Cosmic Dawn refers to the period following the formation of the first stars. During this time, Lyman-$\alpha$ radiation from these stars suffused the universe. Subsequently, the predominantly neutral intergalactic medium became ionized during a watershed known as the Epoch of Reionization (EOR). Numerous projects aim to study these epochs with low frequency radio waves.

Low frequency radio waves probe redshifted cosmic background radiation from this epoch. Lyman-$\alpha$ pumping of cold atomic hydrogen causes an overpopulation of the 21 cm ground state. Level populations fall into equilibrium with the gas temperature because the Lyman-$\alpha$ line profiles are shaped by the thermal motions of atoms. Absorption follows from the radiation temperature cooling less quickly than gas temperature in an expanding adiabatic system. Thus absorption at restframe 21cm wavelength caused by atomic hydrogen in the presence of UV light from the first stars, known as the Wouthuysen-Field effect\cite{1959ApJ...129..536F,1952AJ.....57R..31W} is expected to create an absorption feature. Similar effects may lead to 21cm radiation alternating between absorption and emission as the early universe evolves.\cite{2006PhR...433..181F} Such features are redshifted into the observed FM radio frequency band. 

The Murchison Wide-Field Array (MWA)\cite{2009IEEEP..97.1497L} as well as the Precision Array for Probing the Epoch of Reionization (PAPER)\cite{2013AAS...22110802P} and the subsequent Hydrogen Epoch of Reionization Array (HERA)\cite{2016AAS...22722304P} each received NSF support. They were designed to study the power spectrum of fluctuations in the brightness temperature of 21cm emission from hydrogen. HERA will probe the universe out to redshifts, $z \lesssim 12$. 

The Large Aperture Experiment to Detect the Dark Age (LEDA) seeks to study the sky-averaged 21cm signal and power spectrum from primordial hydrogen out to redshifts $z\lesssim 20$ with an interferometer-based approach\cite{2012arXiv1201.1700G,2018MNRAS.478.4193P,2019AJ....158...84E}. The advantage of this approach over single dipole antennas (discussed below) is the potential for joint estimation of the instrument calibration and a sky model to better control systematic uncertainties. LEDA received several NSF awards beginning with an ATI grant in 2011 (1106059/Greenhill), and it is one of several efforts pursuing this sky-averaged signal.
 
\begin{figure}
   \begin{center}
   \begin{tabular}{c}
   \includegraphics[height=7cm]{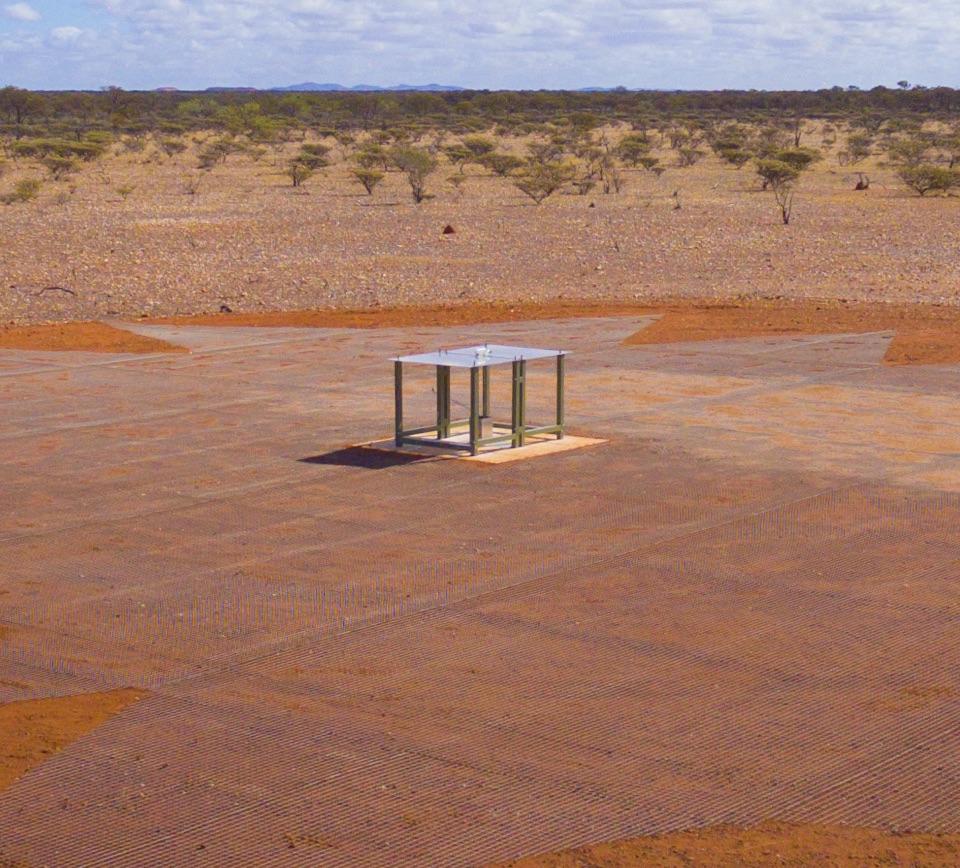}
   \end{tabular}
   \end{center}
   \caption[EDGESImage] 
   { \label{fig:EDGESImage} 
EDGES low-band antenna and ground plane located at the Murchison Radio-astronomy Observatory in Western Australia. Image courtesy of CSIRO.}
   \end{figure}

Notable among these efforts is the Experiment to Detect the Global EOR Signature (EDGES). The EDGES experiment consists of a table-top sized, single dipole antenna that was deployed at the Murchison Radio-astronomy Observatory in Western Australia, see Figure \ref{fig:EDGESImage}. EDGES received three grants from ATI; the initial experiment placed constraints on the abruptness of the EOR transition\cite{2010Natur.468..796B}. Subsequent work improved the sensitivity and calibration of the instrument.\cite{2017ApJ...835...49M}
%
%

In 2018, EDGES reported the first evidence for detection of the 21cm absorption signal corresponding to the birth of the first stars.\cite{2018Natur.555...67B} This result earned widespread attention in the scientific community and the popular media. The public announcement of the scientific paper was aided by a six minute popular science video about the project that was produced by NSF.\cite{FirstStarsVideo} Physics World magazine later hailed the discovery as a ``top ten breakthrough of the year." 

A companion paper proposed to explain the unexpected size of the EDGES signal (a factor of two greater than expected) through interactions between dark matter and baryonic matter.\cite{2018Natur.555...71B} The EDGES signal was also notable for its unexpected width in redshift and its flat profile shape. These unexpected features generated considerable dialogue in the scientific community; questions about the analysis and interpretation of the observations were raised\cite{2018Natur.564E..32H} and rebutted\cite{2018Natur.564E..35B}. Polarized foreground emission may have contaminated the observations\cite{2019MNRAS.tmp.2094S}; however independent analyses of the data are unable to refute the principal claims.\cite{2019ApJ...880...26S}. There is consensus that the EDGES results need confirmation.\cite{2018Natur.555...38G} The EDGES team received a third ATI award in 2019 to continue their work (1908933/Bowman).

\subsection{The Event Horizon Telescope}

\begin{figure}
   \begin{center}
   \begin{tabular}{c}
   \includegraphics[height=7cm]{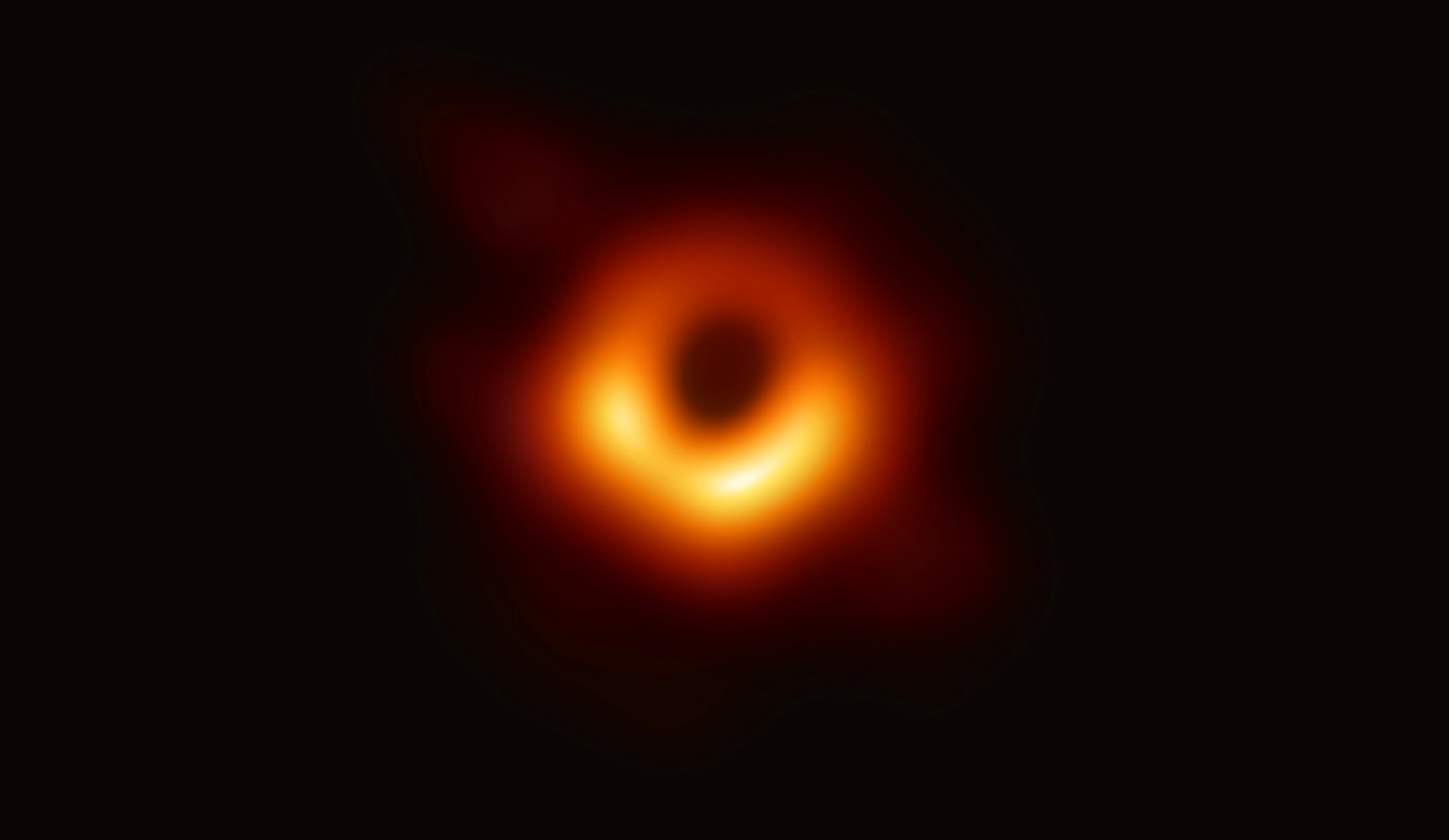}
   \end{tabular}
   \end{center}
   \caption[EHTImage] 
   { \label{fig:EHTImage} 
Image of the black hole at the center of M87 released on April 10, 2019 by the Event Horizon Telescope Collaboration. Image courtesy of Event Horizon Telescope Collaboration.}
   \end{figure}

The Event Horizon Telescope (EHT) has been a remarkable success of the ATI program, and NSF support in general. On April 10, 2019, the EHT collaboration announced the first-ever direct image of a black hole at the center of the galaxy M87, and published six papers documenting their methods and findings\cite{2019ApJ...875L...1E,2019ApJ...875L...2E,2019ApJ...875L...3E,2019ApJ...875L...4E,2019ApJ...875L...5E,2019ApJ...875L...6E}. The public announcement itself consisted of an unprecedented six coordinated press conferences in Washington DC, Brussels, Santiago, Taipei, Tokyo and Shanghai. The NSF played a lead role in organizing and coordinating these events. Resulting media exposure is estimated to have reached {\it billions} of people.\cite{2019CAPJ...26...11C} The EHT image of a black hole, shown in Figure~\ref{fig:EHTImage}, became a popular culture phenomenon, and it was the subject of a dedicated Hearing of the U.S. House of Representatives Committee on Science, Space and Technology.\cite{HouseCommitteeHearingVideo} The EHT collaboration won the Breakthrough Prize for Fundamental Physics in 2020.

%
The EHT image of a black hole in M87 was the culmination of decades of research and development in Very Long Baseline Interferometry (VLBI). VLBI is a technique that synchronizes telescope facilities around the world to effectively synthesize an Earth-sized telescope. Radio wavelength VLBI began in the late 1960's and advanced as new facilities were constructed. Developments were supported by NSF for astronomy and subsequently by NASA and other agencies for geodesy. Beginning in 1975, VLBI experiments at progressively shorter wavelengths sought to constrain the size of the black hole at the center of the Milky Way, Sgr A*, by showing that the source was dominated by scattering (apparent size dependent on wavelength squared). From 1994-2003 the millimeter wavelength VLBI program at MIT Haystack Observatory under Alan Rogers pursued exploratory observations, supported by several NSF awards. 

ATI award 0352953/Doeleman continued this research and accomplished giga-bit per second (Gb/s) recording bandwidths for the first time, which along with the wideband digital backends developed with support from 0521233/Whitney, was a critical step in getting the first 1.3 mm fringes on Sgr A*.\cite{LonsdalePrivateCommunication} It lead to the discovery of event horizon scale structure in Sgr A*\cite{2008Natur.455...78D}. ATI Award 0905844/Doeleman continued development of high speed data capture, demonstrated sustained data recording rate of 16 Gb/s and resulted in detection of horizon scale structure in the galaxy M87\cite{2012Sci...338..355D}. This marked the beginning of a larger push to re-instrument the VLBA. 1207704/Doeleman first deployed 1.3 mm VLBI at the South Pole Telescope, enabling the longest possible baselines to greatly enhance VLBI coverage. 1310896/Doeleman increased the millimeter wavelength VLBI recording capacity to 32 Gb/s, more than an order of magnitude greater than normal VLBA recording rates.\cite{DoelemanPrivateCommunication}

The international EHT collaboration grew from early research collaborations circa 2009 that were formalized at a meeting in Waterloo in 2014. The collaboration received support from the European Research Council, other international agencies and private funders. Some participants provided extensive support out of institutional budgets and the project earned 22 separate NSF awards between 2000-2018.  

The 2017 EHT observing campaign included eight sites in the inter-continental array, and notably included a phased ALMA for the first time. The large collecting area of ALMA allowed sufficient sensitivity to enable successful fringes and finally imaging of the black hole at the center of M87. Observations and analysis, especially of Sgr A*, are ongoing. EHT Founding Director Doeleman acknowledged the importance of early NSF funding to the project in congressional testimony: ``NSF support was crucial. NSF support enabled the small EHT team to grow and carry out key proof of concept experiments."\cite{EHT-C-span}

\section{CONCLUSION}
\label{ConclusionSection}

Over the past thirty years, ATI has supported new technologies and instrumentation for astronomy. The program has been an incubator for transformative research across the entire spectrum of ground-based astronomy, from low-frequency radio waves through optical wavelengths. Typically, ATI has supported projects that are small enough to be managed by a single investigator, often an early career researcher, yet large enough to have a substantial impact. Sometimes the greatest impact is only fully appreciated years or even decades after the award ends. These impacts may be scientific or technical, but they may also be expressed in the professional activities of the investigators themselves. ATI awards have provided important training opportunities and have launched careers.

Some notable firsts that were supported wholly or in part by ATI awards include: the first laser guide star adaptive optics system for astronomy, the first multiconjugate adaptive optics system for solar imaging, the first on-sky demonstration of extreme adaptive optics, the first demonstration of high resolution spectroscopy for exoplanet studies, the first microwave kinetic inductance detector for optical-IR astronomy, the first IR multi-object spectrograph, the first phased array feed for radio astronomy, the first measurements of cosmic microwave background angular power spectrum and polarization, the first evidence of light from the first stars and the first image of a black hole.

ATI has impacted whole fields of research. Examples include optical astronomy, adaptive optics, exoplanets, IR astronomy, millimeter-wave astronomy -- including interferometry, very long baseline interferometry, and bolometers for CMB research, digital signal processing for radio astronomy and low-frequency radio investigations of the epoch of reionization. 

In each of these cases, ATI provided one or more impactful awards that shaped the entire field. Sometimes the impact has been a technological breakthrough that paved the way for others, such as in CCDs, adaptive optics, high-precision radial velocity studies, IR multi-object spectroscopy, and semiconductor bolometers. Sometimes the impact has been to disseminate enabling technology, methods and training to a broad community, such as CCDs, adaptive optics and digital signal processing.  In several instances, ATI awards provided crucial early seed funding for what ultimately became major facilities, such as the Richard F. Caris Mirror Lab at the University of Arizona and the Vera C. Rubin Observatory (formerly LSST). ATI has provided crucial early support to enable a small team to grow toward a larger collaboration, such as the Event Horizon Telescope. In some instances, ATI wholly or substantially supported key scientific discoveries, such as evidence of light from the first stars.

Literature acknowledgement and citation statistics provide a limited means to assess the impact of ATI awards.  Such analysis shows that ATI awards are acknowledged and the resulting papers are cited in the literature at comparable rates to a similar pure science program that has no technology or instrumentation component.  ATI supported investigators write papers to the same degree, and with the same impact as their peers who do not build instruments. Thus the direct impact of ATI awards in the literature is comparable to pure science awards. By considering only direct acknowledgements and only peer-reviewed literature, this conclusion is conservative and robust. The impact distribution of research from both NSF-funded grant programs exceeds that of the general astronomical literature, undoubtedly due to the peer-review selection process of NSF grants. 

However, the total impact of ATI awards cannot be captured by literature acknowledgement and citation statistics alone, as many examples above illustrate. New technology is a science multiplier that enables new fields of study and ways of observing that were never before possible. An award may occur at a pivotal time in a larger project, and as a result its impact may only become apparent in hindsight.  Furthermore, technology matures over a longer time period than an individual award.  Several examples show that transformative technologies for astronomy typically take at least a decade to materialize. Then they open entire new frontiers to science.  Finally, ATI provides opportunities for education and specialized training.  By providing awards that are large enough to make an impact, and small enough for an early career investigator to manage, ATI fills an important niche in the training of instrument developers.  Taking the long view illuminates the tremendous impact of technology and instrumentation for astronomy.

\acknowledgments     
 
For discussions about past awards and/or for reviewing portions of this manuscript, we thank J. Roger P. Angel, Christoph Baranec, Richard Barvainis, Jamie Bock, Judd Bowman, Mark Chun, Scott Diddams, Sheperd Doeleman, Steve Eikenberry, Don Figer, Debra Fischer, Jian Ge, Lincoln Greenhill, Don Hall, Shaul Hanany, Buell Jannuzi, Ed Kibblewhite, Casey Law, Colin Lonsdale, Jared Males, Ben Mazin, James E. Neff, Joseph E. Pesce, Anthony Readhead, Deqing Ren, Ray Sharples, Theo A. ten Brummelaar, J. Anthony Tyson, Wayne van Citters and Jonathan Weintroub.  We also thank Dennis Crabtree for providing data on the median citation rate of publications in {\it The Astronomical Journal}. This research uses Web of Science data by Clarivate Analytics provided by the Indiana University Network Science Institute and the Cyberinfrastructure for Network Science Center at Indiana University. This research has made use of NASA Astrophysics Data System. This research has made use of a Python module to interact with ADS developed by Andy Casey that is available at {\tt  https://github.com/andycasey/ads}. For administrative support, we thank Renee Adonteng and Allison Farrow.


\bibliography{ATIReviewBibliography}   
\bibliographystyle{spiebib}   
%
%
\end{spacing}{2}   
\end{document}